\PassOptionsToPackage{table,xcdraw}{xcolor}

\documentclass[final,5p,times]{elsarticle}

\usepackage{amssymb}

\usepackage{array}
\newcolumntype{P}[1]{>{\raggedright\arraybackslash}p{#1}}
\usepackage{booktabs} 
\newcolumntype{L}[1]{>{\raggedright\let\newline\\\arraybackslash\hspace{0pt}}m{#1}}

\usepackage[hyphens]{url}
\usepackage[hidelinks]{hyperref}
\hypersetup{breaklinks=true}
\usepackage{xcolor}
\usepackage[utf8]{inputenc}
\usepackage[T1]{fontenc}
\usepackage{longtable}
\definecolor{myblue}{RGB}{222,234,246}
\usepackage{caption}

\newcolumntype{T}[1]{>{\centering\arraybackslash}p{#1}}
\usepackage[parfill]{parskip}    

\journal{Journal of Systems and Software}

\begin{document}

\begin{frontmatter}



\title{High-Availability Clusters: A Taxonomy, Survey, and Future Directions}


\author[1]{Premathas Somasekaram}
\author[1]{Radu Calinescu}
\author[2]{Rajkumar Buyya}



 \affiliation[1]{organization={Department of Computer Science, University of York},
            addressline={Deramore Lane}, 
            city={York},
            postcode={YO10 5GH}, 
             country={UK}
}


 \affiliation[2]{organization={Cloud Computing and Distributed Systems (CLOUDS) Lab, School of Computing and Information Systems, The University of Melbourne},
            country={Australia}
}

\begin{abstract}
The delivery of key services in domains ranging from finance and manufacturing to healthcare and transportation is underpinned by a rapidly growing number of mission-critical enterprise applications. Ensuring the continuity of these complex applications requires the use of software-managed infrastructures called \emph{high-availability clusters} (HACs). HACs employ sophisticated techniques to monitor the health of key enterprise application layers and of the resources they use, and to seamlessly restart or relocate application components after failures. In this paper, we first describe the manifold uses of HACs to protect essential layers of a critical application and present the architecture of high availability clusters. We then propose a taxonomy that covers all key aspects of HACs---deployment patterns, application areas, types of cluster, topology, cluster management, failure detection and recovery, consistency and integrity, and data synchronisation; and we use this taxonomy to provide a comprehensive survey of the end-to-end software solutions available for the HAC deployment of enterprise applications. Finally, we discuss the limitations and challenges of existing HAC solutions, and we identify opportunities for future research in the area.
\end{abstract}





\end{frontmatter}
\graphicspath{{./images/}}

\section{Introduction}
\label{sec:introduction-taxonomy-survey}

High-availability clusters (HACs), also called \emph{failover clusters}, are software-managed systems that support the reliable execution of complex enterprise applications (EAs) or of their key layers\footnote{We use the term \emph{layer} for a logical component of an application, and the term \emph{tier} to denote a physical structure, as recommended in~\cite{schmidt2006high}.} and components. They play a major role in ensuring the high levels of availability required of today's mission-critical EAs \cite{liu2003availability,schmidt2006high,marcus2003blueprints}. They comprise physical servers, storage, communication and other hardware infrastructure, together with sophisticated HAC-management software. This software is responsible for continuously monitoring the protected EA layers (e.g., the application server and database layers), and for seamlessly mitigating EA component failures through failover or through automatically restarting, repairing or relocating failed components. As such, HACs enable the delivery of essential EA services with minimal downtime. 

A key requirement for HACs is to ensure the continued operation of the single-point of failure (SPOF) components of the protected EA layers. Such SPOF components may include databases, distributed transaction coordinators, software load balancers, and storage. Due to the diverse high availability (HA) needs of mission-critical EAs, HACs must comply with a wide range of additional requirements. These requirements differ significantly from one EA to another. For example, enterprise resource planning (ERP) EAs facilitate transactions (e.g., online transaction processing), and therefore must be deployed on HACs capable of ensuring the atomicity, consistency, isolation and durability (ACID) requirements associated with transaction processing \cite{marcus2003blueprints, critchley2014high}. In contrast, enterprise analytics EAs (e.g., online analytical processing) do not have these strict ACID requirements, since they tend to operate with read-only data, while transactional EAs are typically read-write intensive \cite{mansouri2018data}. Moreover, the focus of the analytical EAs is to manage large data sets in multiple steps (e.g., staging, transformation, processing, and reporting), and this is typically reflected in the architectural layers and the components that are part of such layers \cite{vercellis2011business, demchenko2014defining, hu2014toward}. Hence, the SPOF components of analytical EAs differ from those of transactional EAs. Therefore, the HACs used to protect such solutions vary significantly.

Further HAC requirements arise from the need to monitor and maintain the ``health'' of the cluster itself. An essential monitor called a \emph{heartbeat} \cite{ranade2003shared,vogels1998design} is required to periodically check the health of individual cluster nodes (i.e., servers) so that the appropriate failover procedure can be initiated when node failures are detected. At cluster level, a \emph{quorum} system \cite{critchley2014high, birman2012guide} is needed for scenarios where the cluster ends up divided into cluster partitions that can no longer communicate with each other. In these scenarios, a voting protocol is enacted to select a single partition that will continue to run the EA. In this way, quorum systems prevent the occurrence of a \emph{split-brain} \cite{schmidt2006high, marcus2003blueprints}, i.e.\ a situation in which multiple partitions attempt to use the EA resources at the same time, potentially corrupting important EA data \cite{ranade2003shared}.

This diversity in HAC uses and requirements has led to significant research on the techniques underlying the operation of HACs. At the same time, the ability of HACs to run critical EAs with minimal downtime prompted the development of multiple end-to-end HAC solutions. Our article provides a taxonomy and a survey of this large body of work. The taxonomy clarifies and formalises the often overlapping or conflicting terminology and classifications used by HAC researchers and developers. The survey comprises two parts. The former part, covered when we define the taxonomy, represents an extensive coverage of the HAC research landscape. This part offers insights into the capabilities and limitations of the techniques used to achieve HA for critical EAs. The latter part covers end-to-end HAC solutions, supporting the developers and users of these solutions. 

Addressing HAC limitations and extending the applicability of HACs to new computing paradigms is an active field of research. Recent advances in areas such as machine learning \cite{Murphy:2012:MLP:2380985, gu2009online} and self-adaptive systems \cite{DBLP:journals/tse/CalinescuWGIHK18} provide new avenues for addressing current HAC challenges, while recent 
technologies such as containerisation \cite{endo2016high, li2015leveraging} require the development of new types of HACs. A discussion of these new directions for HAC research and solutions is also provided in the article. This is the first taxonomy and in-depth survey that focus on HACs. While a few previous studies proposed taxonomies for availability in the cloud \cite{ Nabi2016AvailabilityArt, endo2016high} and dependable and secure computing \cite{ avizienis2004basic},  these taxonomies are complementary to our work, as they do not consider essential HAC characteristics such as heartbeat, quorum, topology and type of cluster.

The contributions of our article are organised as follows. 
Section~\ref{sec:architecture-high-availability-clusters} explains how HACs are used to protect different layers of critical EAs, and introduces a generic HAC architecture and key HAC terminology. Section~\ref{sec:taxonomy} presents the HAC taxonomy and the techniques underpinning core HAC operations such as monitoring, heartbeat, quorum, failure detection, and component failover. Section~\ref{sec:survey} uses the taxonomy to survey end-to-end EA HAC solutions available commercially or from open-source projects. Section~\ref{sec:future-directions-taxonomy-survey} discusses HAC limitations, open challenges, and research opportunities. Lastly, Section~\ref{sec:summary-conclusion-taxonomy-survey} concludes the article with a brief summary.


\section{Uses and Architecture of High-Availability Clusters}
\label{sec:architecture-high-availability-clusters}

\subsection{Key Concepts and Terminology}
\label{sec:key-concept-terminology-HAC}

The ISO/IEC 25010 standard defines availability as the `\emph{degree to which a system, product or component is operational and accessible when required for use}' \cite{ISO:25010:2011}. Availability is calculated as the ratio between the time when a system is operational and the total time over which the system was observed. Equivalently, availability can be computed as the ratio between the \emph{mean time between failures}, $\mathit{MTBF}$, and the sum of the mean time between failures and the mean time to recover after failures, i.e., the \emph{mean time to repair}, $\mathit{MTTR}\,$: $\mathit{availability}=\mathit{MTBF}/(\mathit{MTBF}+\mathit{MTTR})$  \cite{Koren:2007:FS:1206164,OConnor2012}.

Component failures lead to \emph{downtime} (i.e., periods when the system is not operational or accessible), and to a decrease in availability. As such, HACs are responsible for reducing both the frequency and the duration of failures, and thus their impact on the availability of the protected EAs. Discharging the first responsibility involves monitoring specific EA components, to identify and resolve \emph{faults} before they lead to \emph{errors}, and errors before they trigger \emph{failures}, i.e., violations of requirements observable to EA users \cite{Koren:2007:FS:1206164}.

A fault can occur in any resource (i.e., atomic component) of an EA, and the critical resources are usually combined into one or several SPOFs (or SPOF groups). If such a resource fails irrecoverably, it will lead to the failure of its associated SPOF as well. When an SPOF fails, it may bring down an entire application. Achieving high availability requires that the SPOFs of an application are entirely or partially eliminated, or masked. 
Consequently, HACs discharge their second responsibility by relocating SPOF-related resources to a secondary server after irrecoverable failures. In this way, they mask the failures of resources, and thus of application SPOFs. 

In this context, HACs employ a threefold strategy for failure management:
\begin{enumerate}
 \item 	HACs avoid EA downtime, even in the presence of failures of individual resources. To achieve this, HACs reinitialise or restart resources after faults and errors (increasing MTBF) and after failures (reducing MTTR).
 \item 	HACs promote the failure management to a resource group level if the failure at a resource level cannot be resolved locally. This leads to a failover of the concerned resource group to another node. A resource group can also be reinitialised on the same node if there are no available secondary nodes. A resource-group failover is faster than a complete system failover. Therefore, the likely outcome is that the failover does not cause downtime.
 \item 	If there are dependencies between the resource groups and after critical failures, a complete system failure may occur. In this event, the complete system is failed over to another node.
\end{enumerate}
In the first scenario, components are restarted, whereas in the other two scenarios components are first stopped and then started, in a specific order determined by their interdependences.

\subsection{Enterprise Application Layers}
\label{sec:enterprise-application-layers}

EAs such as ERPs are transaction-intensive and require stateful communication. Moreover, data consistency and data integrity are vital for such applications. Additionally, modern EAs are highly integrated, which means that data corruption in one application may lead to data corruption in other integrated systems. Therefore, data corruption and data loss must be prevented even when failures occur. To identify and achieve HA holistically for an EA, it needs to be broken down into a set of essential building blocks that are referred to as \emph{layers}. Critchley \cite{critchley2014high} proposes a layered architecture in describing an IT environment. Somasekaram \cite{mastersthesis} suggests a similar approach of separating the layers of an IT solution for outsourcing purposes. 

When all the layers of an EA are identified, an appropriate solution for ensuring the HA of each layer can be devised. Multiple solutions are typically possible for each layer, including the use of a HAC. As such, different EA critical layers can each be protected by a separate HAC. Alternatively, a single HAC can be employed to protect several critical layers of an EA. In either case, any EA layer not protected by HAC(s) may require other types of HA solutions (e.g., redundancy or fault tolerance). In the special case of applications with only one critical layer (e.g., firewalls), HA can be ensured through using a \emph{single-layer HAC}~\cite{ayuso2009demystifying,schmidt2006high,CheckPointSoftwareTechnologiesLtd2018}.

Based on the solutions that can ensure their availability  
\cite{critchley2014high,mastersthesis,Bajohr2008,Zhu2006,Santos2017,Fernandes2014,barroso2009datacenter,Dukaric2013,AmazonWebServicesInc2016,wang2004architectural,wen2020design}, the components of an EA can be organised into the nine layers from Table~\ref{tab:EA-layers-HA-solutions}. For each layer, the table shows the typical role(s) that the layer can play within an EA, the solutions available for ensuring its availability, and whether an \emph{application HAC} (i.e., a multi-layer HAC) is among these solutions---providing protection for the whole layer or only for its client resources. As indicated in this table, an application HAC can protect the \emph{application server}, \emph{application core}, and \emph{database} layers of an EA, as well as the client resources associated with the EA \emph{network} and \emph{storage} layers.\footnote{The network and storage EA layers are part of the EA infrastructure, and present both a server view and a client view. As an example, a storage system in itself is part of the server view, while its individual disks associated with a server or with a virtual machine become part of the client view, and thus need to be protected by the application HAC.} In contrast, a HAC is not typically used to protect the \emph{operating system}, and the \emph{virtual machine} (VM) or \emph{server} layers of the EA (i.e., layers 4--6 from Table~\ref{tab:EA-layers-HA-solutions}), as a failover always involves relocating the application environment to a different VM or server, respectively. The protection of the \emph{data centre} layer is also beyond the scope of a HAC. However, the HAC still needs to monitor critical elements from layers 4--6 in order to identify critical issues such unacceptably high levels of CPU utilisation for a server.

\begin{table*}
\centering
		\caption{Enterprise application (EA) layers with possible high availability (HA) solutions} 
		\label{tab:EA-layers-HA-solutions}
 		\def\tabcolsep{3pt}
    	\sffamily
		\begin{footnotesize}
            \begin{tabular}{p{4mm}p{30mm}p{40mm}p{42mm}p{12mm}}
		\toprule
		\textbf{No} & \textbf{Layer} & \textbf{Typical Role(s) within the EA} & \textbf{Possible HA solution(s)} &  \textbf{HAC$^\dagger$} \\ \midrule
		1 & application server (e.g., web servers \cite{buyya1999high,SASInstituteInc.2017,Oracle2013}) & key tier in multi-tier EAs, e.g., presentation layer & 
		use multiple instances with load balancing \cite{ranade2003shared, OracleCorporation2017b}
		 & optional \\ \midrule
		2 & application core (e.g., ERP central \cite{lyu2016high,OracleCorporation2016b} & coordination of distributed transactions, application servers
		& use application HAC  \cite{OracleCorporation2016b,schmidt2006high, VeritasTechnologiesLLC2017} & yes \\ \midrule
		3 & database (e.g., Oracle, DB2, HANA \cite{schmidt2006high, bartkowski2012high}) & databases to support the main application &  high-availability features provided by database, such as replication and mirroring \cite{schmidt2006high,minhas2013remusdb,VeritasTechnologiesLLC2017} which can be used with application HAC  & yes \\ \hline
		4 & operating system (e.g., Linux, UNIX) & operating environments & redundant server environment & no \\ \hline
		5 & virtual machine (VM) & VM (e.g., virtualisation platform) & VM cluster \cite{cully2008remus} (a HAC can be combined with a VM cluster \cite{VMwareInc.2015})  & no \\ \hline
		6 & server & server hardware & redundant servers and fault tolerance & no \\ \hline
		7 & network (e.g., private, public networks) & local area network (LAN), virtual LAN (VLAN) & redundant network devices, fault tolerance, hardware HACs (e.g., for routers and load balancers) \cite{barroso2009datacenter,PaloAltoNetworksInc.2018,sheghdara2020automatic} & only client resources \\ \hline
		8 & storage (e.g., the different type of storage systems) & storage area network (SAN), NAS, direct attached storage (DAS) \cite{marcus2003blueprints} & redundant devices, fault tolerance, storage HAC \cite{marcus2003blueprints,Zhu2006,saxena2020cloud} & only client resources \\ \hline
		9 & data centre (e.g., essential data centre components) & supporting utilities such as UPS, power distribution unit (PDU) \cite{zhang2021flex}, cloud operating systems~\cite{heimovski2020ft}, and backup infrastructure & redundancy by multiple sites, and redundant data centre equipment, such as UPS and fault tolerance for components, HA for the individual data centre components \cite{barroso2009datacenter,rosendo2020availability} & no \\ \bottomrule
\multicolumn{5}{l}{$^\dagger$High availability of layer can be ensured by an application HAC}
		\end{tabular}
\end{footnotesize}
	\end{table*}

A few research initiatives have addressed the challenges of achieving HA solutions for multiple EA layers from Table~\ref{tab:EA-layers-HA-solutions}. Bajohr et al. \cite{Bajohr2008} have devised an HA framework for Springer Verlag’s Online Conference Service. Their framework combines different solutions for several layers of this multi-tier applications, including an N+M HAC (the terms are explained in the \textit{topology} section of our taxonomy) for application servers (layer~1), and a master-slave configuration for the database (layer~3). Similarly, Sun et al. \cite{Sun2016} present an HA architecture for a multi-tier application in which multiple HA solutions are combined to enable HA for the application. However, most research to date has focused on HACs for single EA layers. For instance, Cheng et al. \cite{Cheng2005} developed an application cluster service (APCS) scheme comprising separate methods that support state recovery and failure management, respectively. APCS assumes that the state of a shared-storage database layer does not change, and therefore focuses on the protection of the application layer of a three-tier architecture. In many other approaches, the layer protected by different types of HA solutions is the database layer \cite{Xiong2016},  as in the case of Riley et al.'s HA cloud for research computing \cite{Riley2017}.  Built using the OpenNebula cloud computing platform, this solution employs an active-active HA MariaDB cluster (layer 3) to support the storage of cloud objects. 

In summary, modern EAs require a combination of HA solutions to achieve the required levels of end-to-end availability. More often than not, the infrastructure components of EAs have their own HA setups, and thus HACs typically focus on ensuring the availability of the actual applications. 

\subsection{HAC Architecture}
\label{sec:HAC-architecture}

\begin{figure*}
	\centering
	\includegraphics[width=\linewidth]{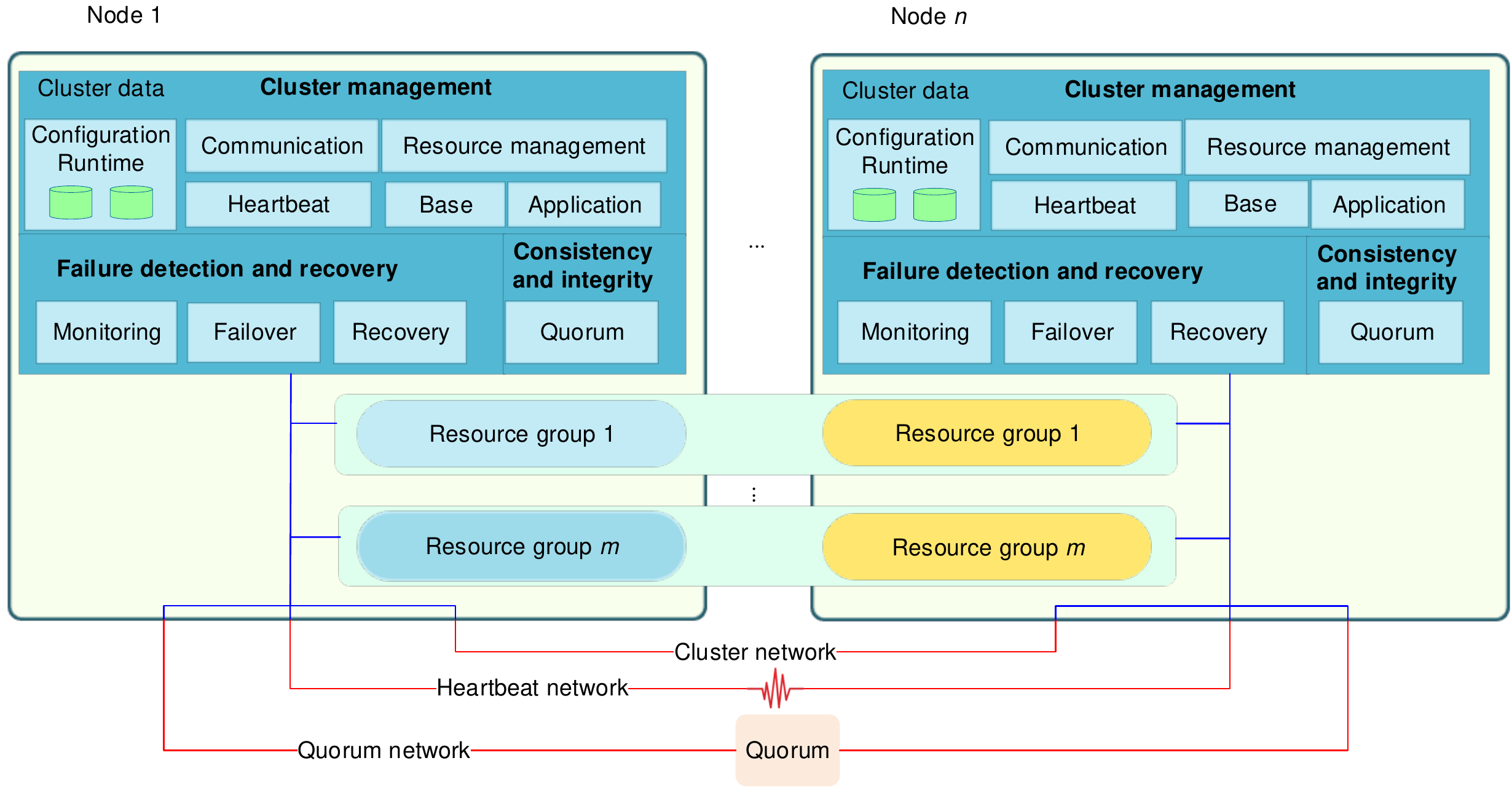}
	\caption{Architecture of a high availability cluster (HAC) with $n\geq 2$ nodes.}
	\label{fig:HAC-architecture}
\end{figure*}

Figure~\ref{fig:HAC-architecture} shows the high-level architecture of a generic HAC operating on $n\geq 2$ nodes distributed across one or multiple locations (i.e., data centres). The HAC is responsible for the management of an EA whose resources are depicted organised into $m\geq 1$ \emph{resource groups}, out of which only the resource groups on the \emph{primary node}~1 are active. The HAC uses three dedicated private networks---a \emph{cluster network} for its communication across cluster nodes, a \emph{quorum network} to connect all nodes to a quorum device (i.e., a facilitator of the quorum service), and a \emph{heartbeat network} whose role is explained later in this section. To HAC modules deployed on each cluster node \cite{ranade2003shared, vogels1998design, leangsuksun2004failure} are described below, using the terminology summarised in Table~\ref{tab:HAC-terminology}.

\begin{table*}
\centering\footnotesize
\caption{HAC terminology}
\label{tab:HAC-terminology}
    	\sffamily
\begin{tabular}{P{1.6cm}p{11.5cm}}

    \toprule
        \textbf{Term} & \textbf{Description}\\
    \midrule
        Resource & A logical or physical component of an EA layer (e.g., an IP address used by a database, or an application component) that is managed as an atomic entity by a HAC, and is either fully operational or unavailable. Resources have interdependencies that can be described by a hierarchical map \cite{ranade2003shared,vogels1998design,marcus2003blueprints}. \\
        Resource group & A set of logically related resources that can be relocated to a secondary node as one entity. As such, each resource can only belong to one resource group. An EA may comprise numerous resource groups, each of which may represent a significant part of the EA, such as a database \cite{schmidt2006high,ranade2003shared, vogels1998design}.\\
        Split-brain & A condition that occurs when a cluster ends up divided into partitions that perform conflicting operations on the same resources, typically causing data corruption \cite{marcus2003blueprints,birman2004adding}. \\
        Amnesia & A condition that occurs when cluster nodes operate with different configurations, e.g., because nodes that are rebooted resume operation with an older configuration. If such nodes are to become primary, a problem is created because they will run with an out-of-date configuration \cite{OracleCorporation2014}. \\
        Switchover & The manual migration of resources from one node to another \cite{critchley2014high}. \\
        Shared storage HAC & A HAC whose cluster members have access to the same storage. However, when it comes to EAs, typically only one node at a time can allocate the shared storage resources, so that data integrity is not affected \cite{ranade2003shared}. \\
        Dependency & Resources and resource groups have dependencies that must be taken into account during a failover and the subsequent restart of services. These dependencies can be modelled using a acyclic directed graph termed a \emph{dependency configuration} \cite{ranade2003shared}.
        \\
    \bottomrule
    \end{tabular}
  \end{table*}

\textbf{I. Cluster management} is the core HAC module, responsible for overseeing the operation of the other modules, and including the following sub-modules:
  \begin{enumerate}
  \item \textbf{Cluster data}, comprising the data stores managed by a cluster, and shared by all nodes.
    \begin{itemize}
\item \textbf{Configuration} data comprises static data (e.g., HAC configuration parameters).
\item \textbf{Runtime} data consists of dynamic data (e.g., current status of the cluster components).
      \end{itemize}
  \item \textbf{Communication}, which manages the communication between the HAC modules on the same node and between the cluster nodes, and the heartbeat communications. 
  \begin{itemize}
   \item \textbf{Cluster} communication (also known as intra-cluster or inter-node communication) deals with communication between cluster nodes.
  \item \textbf{Heartbeat} is an essential health monitor that checks the health of member nodes, notifying the HAC when the heartbeat of a particular node fails \cite{marcus2003blueprints,vogels1998design}. During such an event, the HAC consults with a quorum to ensure that there are enough votes to continue to run the cluster. If it is the active node that has failed, this will result in a failover, provided that the cluster can reach a quorum \cite{ranade2003shared}.
  \item \textbf{Node} communication (also known as intra-node communication) manages communication within a node.
  \end{itemize}
\item \textbf{Resource management}, responsible for managing two main groups of EA resources: 
  \begin{itemize}
  \item \textbf{Base} resources, which include key components such as CPUs and disks.  
    \item \textbf{Application} resources, which are resources specific to the HAC-protected applications.
    \end{itemize}
  \end{enumerate}

\textbf{II. Failure detection and recovery} is the HAC module responsible for managing failovers and recoveries, and includes the following sub-modules:
  \begin{enumerate}
  \item \textbf{Monitoring}, which monitores the EA resources and notifies other HAC components (e.g., resource management) about any problems.
\item \textbf{Failover}, which is responsible for moving resource groups to a secondary node. Depending on the failure type, a failover can be at resource group or system (i.e., complete application) level. The latter involves moving all resource groups that belong to an application \cite{marcus2003blueprints,critchley2014high}.
\item \textbf{Recovery}, which decides whether failures need to be resolved at resource, resource group or node level by considering their criticality and resource dependencies. When a failure cannot be resolved, the failover sub-module is notified, so that failover can be initiated. 
  \end{enumerate}

\textbf{III. Consistency and integrity} is the HAC module that ensures consistency and integrity across all cluster nodes through the following sub-modules:

  \begin{enumerate}
    \item \textbf{Fencing}, which is a protection mechanism that isolates a resource or node that experienced failures, removing its ability to connect to any of the critical EA resources \cite{critchley2014high,SUSELLC2017}.
    \item \textbf{Quorum}, which is a voting system for determining which partition is allowed to run a cluster when a split of the cluster occurs \cite{critchley2014high,vogels1998design, Quintero2013}. The partition that has the quorum is considered quorate, and can be used to run the cluster without causing a split-brain. 
      \end{enumerate}

These terms are described in greater detail in the taxonomy section below.
\section{Taxonomy of High-availability Clusters \label{sec:taxonomy}}

\makeatletter

\begin{figure*}
	\centering
	\includegraphics[width=.95\linewidth]{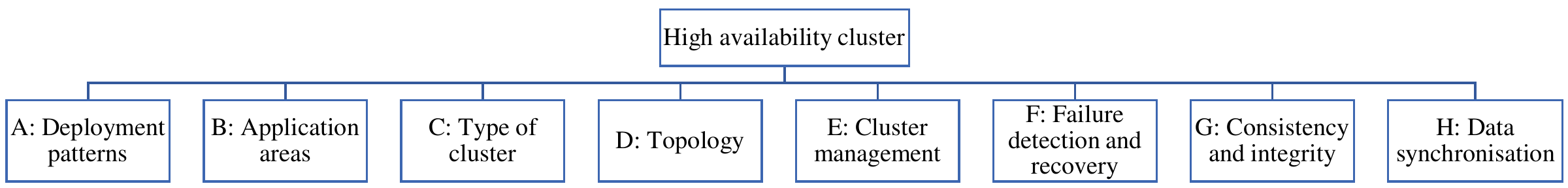}
	\caption{Top-level classes of the HAC taxonomy.}
	\label{fig:top-level-classes-HAC-taxonomy}
\end{figure*}

Our taxonomy applies to single-layer HACs and multi-layer HACs, which it organises into the eight top-level classes shown in Figure~\ref{fig:top-level-classes-HAC-taxonomy}. The first four classes capture how HACs are deployed (\emph{deployment patterns}), which EA layers are protected by HACs (\emph{application areas}), how this protection is achieved (\emph{type of cluster}), and how the HAC nodes are structured and interconnected (\emph{topology}). The next two classes reflect how HACs manage the resources of the protected EA (\emph{cluster management}) and perform detection of and recovery from failures of these resources (\emph{failure detection and recovery}). Finally, the last two classes indicate how the HACs preserve the consistency of the EA data and the integrity of the cluster (\emph{consistency and integrity}), and how the EA data are synchronised across cluster nodes (\emph{data synchronisation}). 

The taxonomy aims to achieve a balance between: (a) considering virtualized resources explicitly where their use makes a significant difference in how a HAC aspect is implemented or operates (e.g., shared-storage or network); and (b) keeping taxonomy classes and subclasses non-prescriptive about the use of virtualised or physical resources where this choice has limited impact (e.g., for monitoring, fencing, heartbeat or quorum).

Given the significant industrial relevance of HACs, we developed the HAC taxonomy based not only on research papers but also on a wide range of technical guidebooks \cite{schmidt2006high,marcus2003blueprints,critchley2014high,Quintero2013,Ranade2002,toeroe2012service,10.5555/2769780}, technical reports \cite{IDCCorporation2016,ServiceAvailabilityForum2011}, white papers \cite{by2006guide,Redhatsap2014,NovellSAP2014,SuseSAP2012}, HAC product documentation \cite{OracleCorporation2016b,OracleCorporation2017a,VeritasTechnologiesLLC2017,VeritasTechnologiesLLC2017a,IBMCorporation2018a,FujitsuLimited2017,MicrosoftCorporation2011} and best practices \cite{Oracle2013,VMwareInc.2015,AzureSAP2020,MicrosoftCorporation2017}.

\subsection{A: Deployment Pattern}

\begin{figure*}
	\centering
	\includegraphics[width=0.6\linewidth]{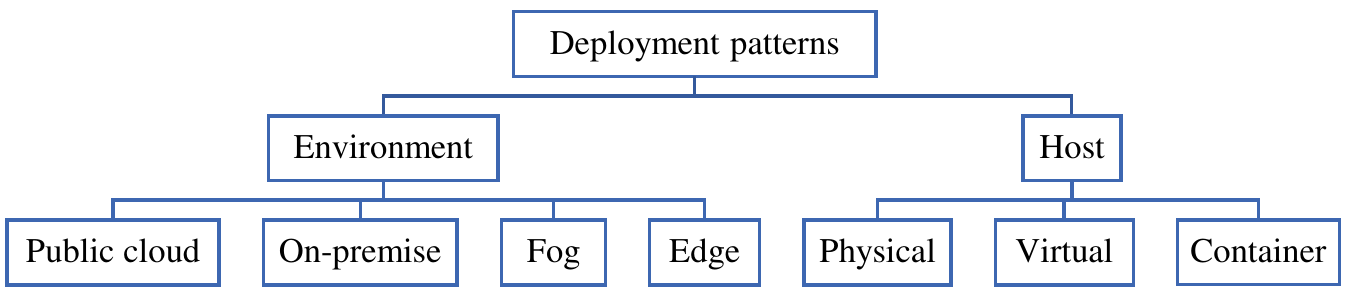}
	\caption{Deployment patterns.}
	\label{fig:deployment-patterns}
\end{figure*}

 The \emph{deployment pattern} of a HAC represents the platform \textbf{where} the HAC solution is deployed. As shown in Figure~\ref{fig:deployment-patterns}, we distinguish between the deployment \emph{environment}---which case be public cloud, or can be on-premise, fog or edge IT infrastructure of the organisation using the HAC, and the type of \emph{host} used for the cluster---which can be physical, virtual or container.

The deployment pattern decides, along with business requirements and technical capabilities, \textbf{what} \textit{cluster type} can be implemented for an \textit{application area}. Table~\ref{tab:deployment-patterns-application-areas} describes the relationship between deployment patterns, application areas, and the rest of the taxonomy. A \textit{cluster type}, on the other hand, decides \textbf{how} an\textit{ application area} can be protected and the subsequent topology and related configuration. An example of this is as follows: if the deployment pattern is a set of virtual servers in a single data centre, it will not be possible to deploy topologies such as \textit{metro} or \textit{continental} (described under \textit{Type of cluster}). Thus each HAC solution comes with deployment restrictions (e.g., whether it can be deployed in a public cloud or not). 

Cloud environments impose restrictions that can cause problems for a HAC because many of the infrastructure elements that a HAC needs to monitor and manage may not be available for a cloud deployment. However, a distinction needs to be made between private (i.e., on-premises) clouds and public clouds because private clouds offer much more flexibility, and functionalities may be identical to an on-premises physical environment. Furthermore, the roles and responsibilities of different stakeholders play an essential role when deploying a HAC in a cloud. Table~\ref{tab:roles-responsibilities-service-models} describes the roles and responsibilities of customers and cloud providers for private clouds and for the service models available in public clouds \cite{liu2011nist,MicrosoftCorporation,AmazonWebServicesInc.2018}, showing that multiple stakeholders may need to collaborate to support the different layers of a HAC in a public cloud.

Both the fog and edge computing paradigms have the potential to improve response times for EAs. However, ensuring HA could be a challenge considering the limited infrastructure components available; some of them are required for implementing HACs (e.g., shared storage) in these deployment environments \cite{yousefpour2019all,qiu2020edge}.  Moreover, the deployment hosts can also change, such as using containers to host an application, in which case a HAC must understand the implementation in order to deliver the required HAC functionalities. A HAC requires continuous monitoring for the protected application and the operating environment to detect failures and resolve them at a granular level. For example, suppose that an application is deployed in a container using the deployment environment edge. The HAC must then ensure that the key resources of the application deployed in the container can be monitored and that the HAC can collaborate with the responsible container orchestration system to ensure that failure mitigation actions can be performed. This requires the HAC and orchestration system to collaborate and support each other. We discuss this in detail in  Section~\ref{sec:analysis-survey-results}.

\begin{table*}
	\centering\footnotesize
	\caption{Connection between deployment patterns, application areas and the rest of the taxonomy}
	\label{tab:deployment-patterns-application-areas}%
	\begin{tabular}{P{.1\textwidth}P{.24\textwidth}P{.24\textwidth}P{.26\textwidth}}
			\toprule
		& \textbf{Deployment Patterns} & \textbf{Application Areas} & \textbf{Rest of the Taxonomy} \\
		\midrule
		Objectives & \textbf{Where} to deploy the solution? & \textbf{What} application or application components need to be protected? & \textbf{How} should the solution be set up to meet the requirements? \\
		& & & \\
		Examples & data centre locations, public cloud, virtual server, container & enterprise system, NAS, network appliance (e.g., firewall), storage system & cluster type, topology, replication, mirroring \\
		\bottomrule
	\end{tabular}%
\end{table*}%

\begin{table*}
	\centering\footnotesize
	\caption{Roles and responsibilities for service models in a public cloud, and for on-premises deployment}
	\label{tab:roles-responsibilities-service-models}%
	\begin{tabular}{p{2em}p{12em}p{6em}p{3em}p{3em}l}
		\toprule
		\textbf{No} & \textbf{Layers} & \textbf{On-premises} & \textbf{IaaS}  & \textbf{PaaS}  & \textbf{SaaS} \\
		\midrule
		1     & Application server  & C     & C     & C     & AP \\
		2     & Application core  & C     & C     & C     & AP \\
		3     & Database  & C     & C     & C     & AP \\
		4     & Operating system  & C     & C     & CP    & CP \\
		5     & Virtual machine  & C     & CP    & CP    & CP \\
		6     & Server  & C     & CP    & CP    & CP \\
		7     & Network  & C     & CP    & CP    & CP \\
		8     & Storage  & C     & CP    & CP    & CP \\
		9     & Data centre  & C     & CP    & CP    & CP \\
		\bottomrule
		\multicolumn{6}{l}{\begin{scriptsize}\textbf{Key:} IaaS -- Infrastructure as a Service; PaaS - Platform as a Service; SaaS -- Software as a Service;\end{scriptsize}}\\
		\multicolumn{6}{l}{\begin{scriptsize}C -- Customer;  AP -- Application provider; CP -- Cloud provider (who may also be application provider\end{scriptsize}}\\
		\multicolumn{6}{l}{\begin{scriptsize}for the SaaS service model SaaS)\end{scriptsize}}\\
	\end{tabular}%
\end{table*}%

\subsection{B: Application Areas}

\begin{figure*}
	\centering
	\includegraphics[width=.99\linewidth]{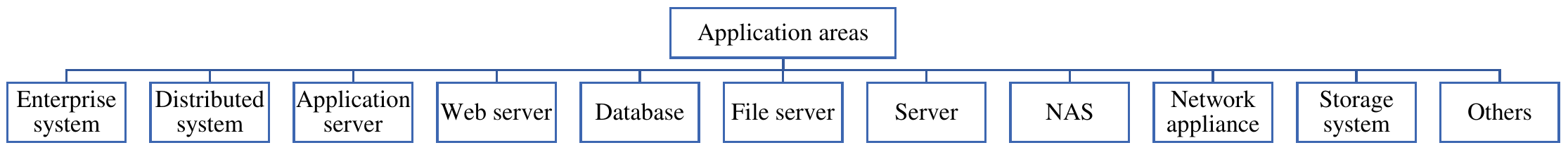}
	\caption{Application areas of HACs.}
	\label{fig:HAC-application-areas}
\end{figure*}

Application areas are the different IT solutions that can be protected by HACs, and a list of typical applications areas is presented in Figure~\ref{fig:HAC-application-areas}. Considering the application area has dual purposes: (1)~to identify if HACs can support the multiple layers that an application is composed of; and (2)~to address all areas that are part of an IT solution, so that HA requirements for those areas can be achieved. For instance, the application area \textit{enterprise system} may require other related areas, such as \textit{application server}, \textit{database}, \textit{server}, \textit{network}, and \textit{storage} to be included to ensure that the \textit{enterprise system} is protected across all critical layers. Some layers can be protected by an application HAC while others may require a different set of options which may include application area specific HACs (presented in Table~\ref{tab:EA-layers-HA-solutions}) \cite{wen2020design}. Moreover, application areas with fewer layers may need to protect fewer components \cite{magnanini2021scalable}. For instance, a HAC in the context of a distributed system (e.g., high-performance computing---HPC) may need to protect fewer components than an EA HAC. In case of an HPC, a head node (principal node) is identified as a SPOF. Thus, a HAC can be deployed to protect the head node \cite{uhlemann2006joshua}. Therefore, the application areas of a solution are determined dynamically during an implementation phase, and the numbers of protected resources will change with the type of primary application to be protected. 

Several recent projects have implemented HACs that support multiple application areas, as also discussed in Section~\ref{sec:enterprise-application-layers}. Xiong et al. \cite{Xiong2016} present a HAC for a relational database in a multi-cloud environment which supports the requirements of both HA and Disaster Recovery (DR). Engelmann et al. \cite{engelmann2008symmetric} have experimented with a HAC to protect the head nodes of an HPC environment. Addressing complex systems that consist of multiple layers, hence also several application areas, is a challenge. Wang et al. \cite{wang2004architectural} address the challenge by proposing an HA solution for a comprehensive medical system which consisted of several layers hence also multiple application areas. The proposed solution used a multitude of HACs to enable HA across the different layers.

\subsection{C: Type of Cluster}
\begin{figure}
	\centering
	\includegraphics[width=.8\linewidth]{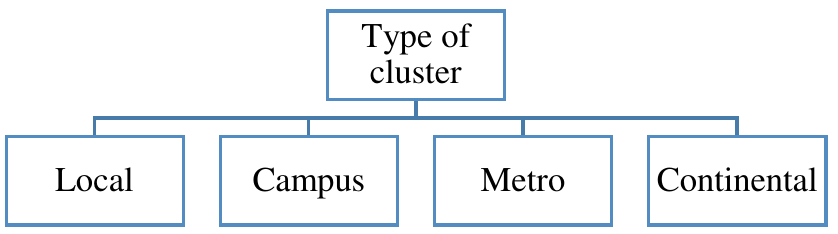}
	\caption{Type of cluster.}
	\label{fig:type-of-cluster}
\end{figure}

The type of cluster plays a vital role in selecting the right topology and related configuration for a HAC. An important characteristic is the distance between nodes, and therefore the number of sites (e.g., data centres). The type could be chosen to meet business requirements, such as business continuity or DR. A DR solution requires at least two data centres with a sufficient distance between them and a related configuration. When a HAC solution is explicitly deployed to support DR, it has to comply with restrictions (e.g., low network latency between data centres). Moreover, supplementary mechanisms must be used to guarantee data integrity during failovers. Therefore, the type of cluster should be treated as the starting point for HAC selection, along with the two top-level taxonomy classes presented previously.  There are four types of clusters, as shown in Figure~\ref{fig:type-of-cluster} and  Table~\ref{tab:types-of-clusters}. Based on a rule of thumb derived from \cite{marcus2003blueprints,schmidt2006high,by2006guide,HewlettPackardEnterpriseDevelopmentL.P.2011,IBMCorporation2017b}, we assumed a communication speed of 3 ms per 160 km to calculate network latency. The distances described in the table may differ due to the use of different technologies. Moreover, the different HAC solutions can also come with specific recommendations.

\begin{table*}
	\centering\footnotesize
	\caption{Type of clusters and potential configurations}
	\label{tab:types-of-clusters}%
	\begin{tabular}{lP{3.5em}P{5.5em}P{4.5em}P{3.8em}P{12em}}
		\toprule
		Type of HAC & Distance (in km) & Network Latency (ms) & Data Centres & Storage Systems & Disaster Recovery Support \\
		\midrule
		Local & $\leq$1    & $\leq$1    & $\geq$1    & $\geq$1    & No \\
		Campus & $\leq$30   & $<$1    & $\geq$1    & $\geq$1    & Limited due to short distance \\
		Metro & $\geq$30   & $<$5    & $\geq$2    & $\geq$2    & Limited due to distance \\
		Continental & $\geq$300  & $>$5    & $\geq$2    & $\geq$2    & Yes \\
		\bottomrule
	\end{tabular}%
\end{table*}%

The limitations presented in Table~\ref{tab:types-of-clusters} are due to the different classes of DR that can be supported for the different types of HAC \cite{schmidt2006high,10.5555/2769780}, which in turn are related to the distance between the data centres. For example, the DR class ``data centre network failure'' can be supported by the campus, metro and continental cluster types. However, a DR class of flooding in the area can only be supported by metro or continental. On the other hand, the continental cluster type can support all DR classes, including the most severe ones, such as an earthquake in the region. Therefore, the most severe DR classes are only supported by the continental cluster type.

\smallskip
\noindent
\textbf{C.1: Local.} A local HAC is hosted in one data centre and uses one storage system, usually shared. When there are two data centres, the distance between data centres is often less than one km \cite{SUSELLC2017}. In such case, there exist two options. Option 1 is to distribute the HAC nodes across two data centres, with all nodes utilising shared storage from one of the data centres. Option 2, on the other hand, uses two storage systems in the two data centres, with the HAC becoming a shared-nothing cluster. However, because data integrity is crucial for EAs, either replication or mirroring must be enabled to synchronise data between the two data centres. The two-data centre setup with replication or mirroring is also a feasible solution for other types of cluster. Since there is usually one data centre associated with a local cluster, the setup is not compliant with DR requirements. 

\smallskip
\noindent
\textbf{C.2: Campus.} A campus cluster is usually deployed across two or more data centres, and the distance between the data centres is less than 30 km \cite{shankar2013enhanced,SUSELLC2017}. Since a campus HAC has a redundant setup for data centres and related components, it can comply with DR requirements (e.g., it can handle DR scenarios such as a data centre failure). However, the distance requirement between data centres means that businesses may opt for other types of HACs which are optimised for longer distances. Nevertheless, campus clusters can support longer distances when combined with other HAC types, becoming \emph{hybrid} clusters — for instance, multiple interconnected campus clusters with one campus cluster functioning as the primary. This setup enables failover locally for most incidents but will trigger a failover to a different site only when a DR scenario takes place at the primary site.

\smallskip
\noindent
\textbf{C.3: Metro.} In a metro cluster, the nodes are distributed across a distance of up to 300 km. Although there is no definite cut-off for this distance, the restrictions come from the techniques that are employed to synchronise data \cite{IBMCorporation2017b,OracleCorporation2016a}. For example, in some cases, the distance can be extended to 400 km by employing Wave Division Multiplexors (WDM) \cite{OracleCorporation}.

\smallskip
\noindent
\textbf{C.4: Continental.} When cluster nodes are geographically dispersed, usually at a distance of more than 300 km, the cluster is characterised as a continental cluster. \cite{HewlettPackardEnterpriseDevelopmentL.P.2012}. A continental cluster can also be referred to as a global cluster or geo-cluster.

\subsection{D: Topology} 
\begin{figure*}
	\centering
	\includegraphics[width=.7\linewidth]{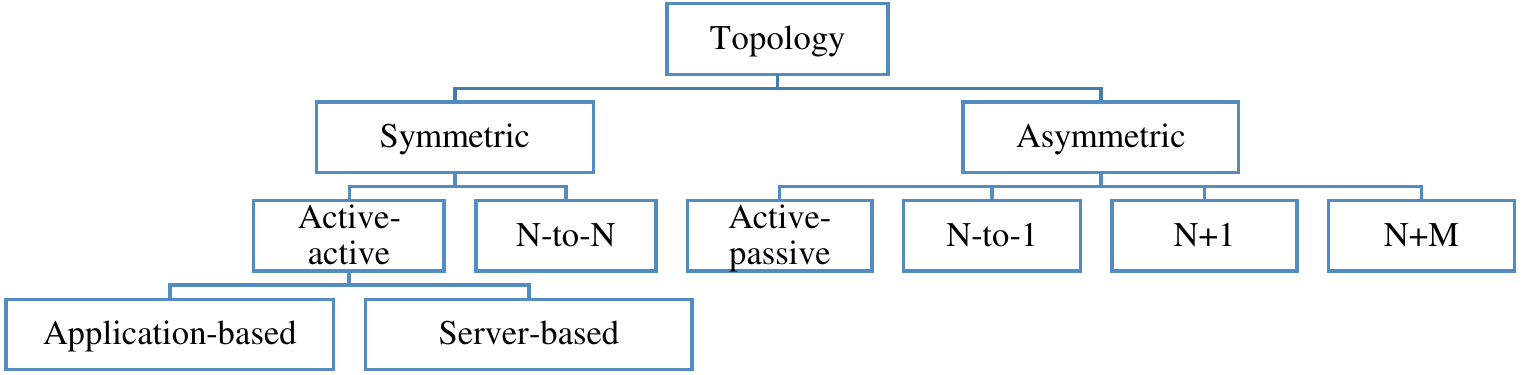}
	\caption{HAC topology.}
	\label{fig:HAC-topology}
\end{figure*}

The topology (or \textit{redundancy model} \cite{kanso2014comparing}) of a HAC represents the way in which the HAC nodes are structured and linked. The topology of a HAC  (Figure~\ref{fig:HAC-topology}) depends on multiple characteristics of its nodes, on the roles of these nodes (primary or secondary), and on its communication devices, networks, storage systems, and supporting tools (e.g., quorum devices).

\smallskip
\noindent
\textbf{D.1: Symmetric.} 
In a \emph{symmetric} 
topology, all cluster nodes can be utilised concurrently: there is no standby node. 

\smallskip
\noindent
\emph{D.1.1: Active-active.} While symmetric active-active describes that all nodes are utilised, there have been research efforts to implement variations of the topology to address the specific needs of distributed systems. Engelmann  et al. \cite{engelmann08symmetric3,he2009symmetric} implemented a prototype with a symmetric active-active topology that operated on more than two nodes to provide HA for an HPC.  The prototype employed two replication mechanisms, internal and external, using reliable and totally ordered message delivery. The internal replication provided synchronisation for the HPC file system metadata service, while the external replication supported the same for the HPC job and resource manager \cite{he2009symmetric}. The evaluation of the prototype showed that the availability could be improved significantly as more nodes were added to the cluster. Therefore, depending on how applications are hosted on such HACs, we distinguish between \textit{symmetric application-based} and \textit{symmetric server-based} topologies. 

\smallskip
\noindent
\emph{D.1.1.1: Application-based.} In a symmetric application-based topology, an application is active on all available cluster nodes. This topology requires application support because managing transactions across multiple nodes is only possible by using additional mechanisms, such as distributed lock management. A component of an application, for instance, a database, may provide these mechanisms, which can then be combined with a HAC solution \cite{OracleCorporation2017a}. For example, IBM Purescale supports parallel access to IBM DB2 databases \cite{bartkowski2012high}, and Oracle provides active-active concurrent access support for Oracle databases using the Oracle Real Application Clusters \cite{Sun2016,VeritasTechnologiesLLC2017a,OracleCorporation2016}.

\smallskip
\noindent
\emph{D.1.1.2: Server-based.} A symmetric server-based topology is frequently referred to as an active-active topology, and this implies that multiple applications are hosted on all server nodes of a cluster; hence, the servers are fully utilised \cite{VeritasTechnologiesLLC2017}. Since all servers are utilised, the topology is considered active-active. When a failover takes place for one or more applications, they failover to one or more of the available servers, implying that a standby node is not required.

\smallskip
\noindent
\emph{D.1.2: N-to-N.} In the \emph{symmetric N-to-N} topology, multiple applications share the same set of N servers, like for the symmetric server-based topology. Upon failure of a primary node for an application, the application is failed over to one of the predefined member nodes of the cluster \cite{distefano2010availability}. The new server will then host both the application that has failed over, and the previously running application \cite{VeritasTechnologiesLLC2017}. The topology supports failing over multiple applications to multiple nodes.

\smallskip
\noindent
\textbf{D.2: Asymmetric.} An asymmetric topology is an active-passive configuration in which one node is active while one or more nodes are in a passive or a standby mode \cite{bouizem2020active}. 

\smallskip
\noindent
\emph{D.2.1: Active-passive.} An active-passive topology is the typical asymmetric topology consisting of a two-node cluster setup in which one node is active while the other node is passive or standby. This topology is sometimes referred to as 2N redundancy \cite{ServiceAvailabilityForum2011}. Today's HACs make a distinction between the different layers of an application. In protecting a layer~3 component (i.e., database), a HAC can either manage it by employing a database-specific extension (agent) or utilising replica or mirroring features that are offered natively by the database \cite{10.1145/3442197}. Most database vendors provide a replica or mirroring option to set up standby databases of primary databases~ \cite{pohanka2020evaluation}, and this configuration can effectively be integrated with a HAC. The prerequisite in such a case is that the HAC has support for the specific feature so that the HAC can recognise and support it as part of its operations. 

\begin{table*}
	\centering\footnotesize
	\caption{Active-passive topology variants}
	\label{tab:active-passive-topology-variants}%
\begin{tabular}{P{4em}P{4.2em}P{8.8em}P{25em}}
		\toprule
		Standby Mode & Recovery Time  & Data Synchronisation Method & Description \\
		\midrule
		\textbf{Active-Cold} & Hours & Backup/restore & A secondary node is installed and configured but brought up only when the primary node is down. Subsequently, the related services are started, as well \cite{levitin2014cold}. \medskip \\
		\textbf{Active-Warm} & Minutes & Mirroring, shared storage & The secondary node is installed and configured and is running. Related application services are started upon failure of the primary node. \medskip \\
		\textbf{Active-Hot} & Seconds & Mirroring, replication, shared storage & The secondary node is fully installed and configured, and services are also started. The secondary node takes over responsibilities immediately upon failure of the primary node. \\
		\bottomrule
	\end{tabular}%
\end{table*}%

Several variants of the active-passive topology exist, depending on the set up for a standby database and for the secondary node \cite{bartkowski2012high,critchley2014high,schmidt2006high}, as shown in Table~\ref{tab:active-passive-topology-variants}. While the standby modes from this table are often used with databases, other application layers may also employ a similar configuration. For example, a layer two component (i.e., application core), employs the active-warm (or warm) standby mode due to the limited need for data synchronisation. However, a common setup by a HAC is to employ either active-hot (or hot) standby or active-warm because otherwise failover time and MTTR will increase and, as a consequence, availability will go down. The primary reason for using the active-cold (or cold) standby or active-warm standby is cost, as using an active host node is associated with higher costs. The standby modes are usually not explicitly supported by modern HAC solutions; instead, the different standby modes of databases and related features that are supported are specified \cite{bartkowski2012high,schmidt2006high}. Moreover, the standby modes are frequently used to refer to the modes of the data centres, particularly in the context of establishing DR for a system \cite{nguyen2016availability}.

\smallskip
\noindent
\emph{D.2.2: N-to-1.} In an N-to-1 topology, multiple applications are supported by one dedicated standby node. Hence the name N-to-1 \cite{distefano2010availability,VeritasTechnologiesLLC2017}. If a node fails, the application is failed over to the standby node and made available there temporarily. However, while the application is active on the standby node, there will be no HA for that application until the primary node is back online. Another aspect of an N-to-1 topology is that such a standby node must be able to host all N applications simultaneously. Hence, sufficient capacity must be available on the standby node.

\smallskip
\noindent
\emph{D.2.3: N+1.} In an N+1 topology, one passive (spare) node supports multiple active applications, similarly to the N-to-1 topology. However, unlike the N-to-1 topology, the N+1 topology employs a \emph{rotation scheme} for failovers \cite{VeritasTechnologiesLLC2017}. This means that, during a failover, an application is failed over to the standby node, but the failed node, once the problems are resolved, effectively becomes the standby node. Hence, any node in the cluster can become a standby node. A variant of the N+1 topology that uses 2+1 nodes (with two active nodes and one node operating as a standby or backup) has been referred to as \emph{asymmetric active-active} in the context of HACs for HPC \cite{leangsuksun2005asymmetric}.

\smallskip
\noindent
\emph{D.2.4: N+M.} The N+M topology refers to HACs that comprise N active nodes and M passive nodes in the cluster, and is called an \emph{N+N} topology when the number of passive nodes equals the number of active nodes.  The topology is employed when one passive node is not sufficient, and $\textrm{M}\!>\!1$ passive nodes are required for failovers \cite{kanso2014comparing,gonccalves2020resource}.

\subsection{E: Cluster Management}
\begin{figure*}
	\centering
	\includegraphics[width=0.82\linewidth]{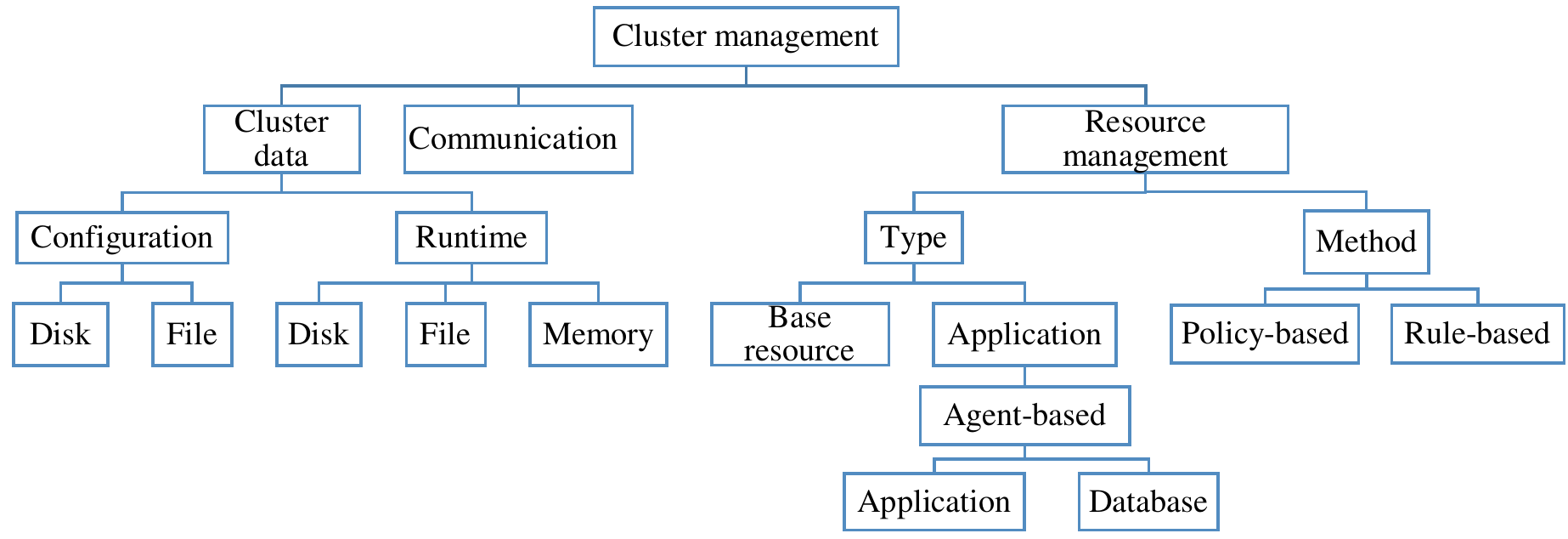}
	\caption{Cluster management.}
	\label{fig:cluster-management}
\end{figure*}

The cluster management module of a HAC is responsible for managing the resources, resource groups, nodes, heartbeats, cluster data, and failovers of a cluster, directly or through other modules. The characteristics used to distinguish between different types of HAC cluster management are shown in Figure~\ref{fig:cluster-management} and described below.

\smallskip
\noindent
\textbf{E.1: Cluster data.} Two types of cluster data are relevant to HACs: configuration and runtime. \emph{Configuration} data contains configuration details of a HAC while \emph{runtime} data stores status of the cluster components. Cluster data can be stored in three types of repositories: disk, file and memory. 

A repository can be either local or shared. However, a prerequisite for a repository is that it is accessible by all cluster nodes. Hence, if a repository is local, a replication mechanism is used to replicate it between the nodes at regular intervals. However, in some cases, when persistent files are employed, the replication is a manual activity. Both in-memory and file repositories are local. However, there are differences in what cluster data type they support. Configuration data is static and is commonly stored in files, while an in-memory repository is generally used to store runtime data to capture changes in real-time. This means there is a rigorous requirement for in-memory repositories to replicate data to other nodes. Therefore, a designated process governs the synchronisation of the runtime data, for instance, Designated Coordinator (DC) in the case of a Pacemaker-based HAC \cite{vogels1998design,SUSELLC2017}. The coordinator ensures that one master repository exists in the primary node while a copy of it, a replica, is distributed across all the member nodes. A shared repository (e.g., disk or file share), on the other hand, stores both configuration and runtime data.
In many cases, a quorum repository, which is shared, can support the requirements. Cluster data in a repository is organised using an information model. For instance, a Cluster Information Base (CIB) uses an XML- based object model to represent both configuration and runtime data. However, there are no standardised information models for dealing with cluster data, and, as such, HACs use different information models. Open Service Availability Framework (OpenSAF), for example, employs Information Model Management (IMM), and objects represent the two types of data: configuration and runtime \cite{toeroe2012service}.

\begin{figure*}
	\centering
	\includegraphics[width=\linewidth]{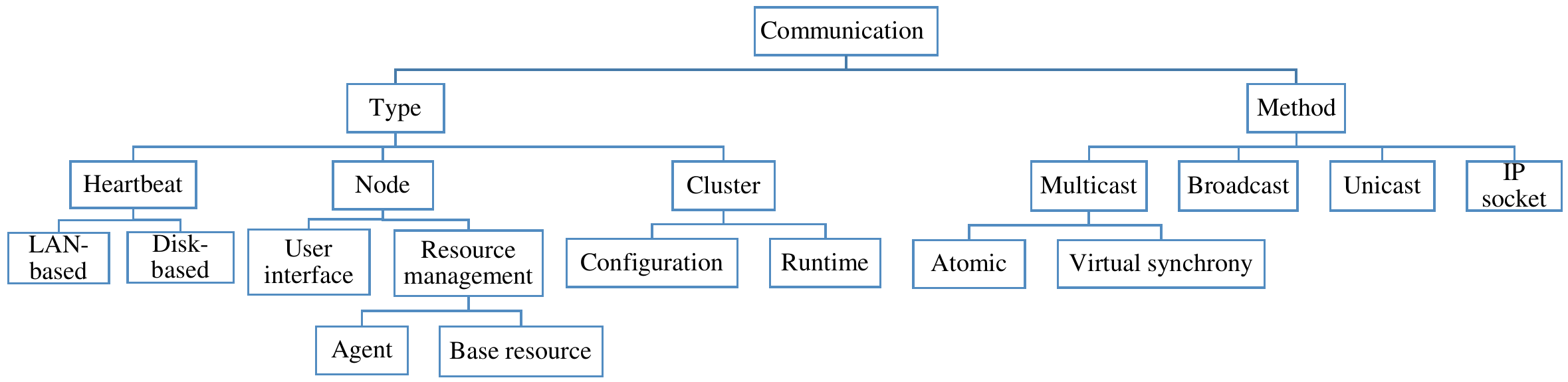}
	\caption{Cluster communication.}
	\label{fig:cluster-communication}
	\end{figure*}

\smallskip
\noindent
\textbf{E.2: Communication.} HACs can use different communication types and methods (Figure~\ref{fig:cluster-communication}).

\smallskip
\noindent
\emph{E.2.1: Type.} A communication type describes the different kinds of communications that a HAC employs and is split further into three subclasses: heartbeat, node and cluster. 

\smallskip
\noindent
\emph{E.2.1.1: Heartbeat.} A heartbeat is a form of intra-cluster communication. However, it is separated in the taxonomy to highlight its importance and use of additional resources, such as a dedicated network. The type and content of heartbeat messages differ from solution to solution. In some cases, a heartbeat message could be a simple ping or a keepalive to provide the status of a cluster node \cite{hou2003design,bouizem2020active}.  Heartbeat communication use a LAN-based or a disk-based method \cite{marcus2003blueprints,VeritasTechnologiesLLC2017,SUSELLC2017,IBMCorporation2016a}.

\smallskip
\noindent
\emph{E.2.1.1.1: LAN-based} heartbeat communication uses a Transmission Control Protocol/Internet Protocol (TCP/IP) network \cite{marcus2003blueprints,schmidt2006high}. Since heartbeat is a key component of a HAC, the recommendation for business-critical solutions is to set up a dedicated network, such as a virtual LAN, to facilitate heartbeat communication \cite{marcus2003blueprints,VeritasTechnologiesLLC2017,IBMCorporation2016a}. With this approach, the heartbeat traffic is not disturbed or delayed by other kinds of traffic in a network, which could be the case if the network is shared. Furthermore, adding redundancy to a heartbeat network by using multiple networks is also a good option so that a single network does not become a SPOF. 

\smallskip
\noindent
\emph{E.2.1.1.2: Disk-based} heartbeat uses a shared disk and also the SAN fabric as a means to facilitate communication \cite{IBMCorporation2016a}.
In some cases, LAN-based and disk-based heartbeat types can be combined to create a full heartbeat service. If a heartbeat mechanism is not employed, an alternative and robust mechanism is required to detect node failures. Cheng et al. \cite{Cheng2005} propose a HA solution that employs a module that can detect whether a node is sick or not and subsequently forecast the time of failure. This renders the heartbeat setup to be nonessential in such cases.  However, there is no information regarding how such a solution works when there are many nodes in a cluster.

\noindent
\emph{E.2.1.2: Node} communication is referred to as \emph{intra-node} and deals with communication within a cluster node. The node communication uses internal communication schemes, for instance, inter-process communication (IPC) within a server. Two types of such communication exist in HACs: user interface and resource management.

\smallskip
\noindent
\emph{E.2.1.2.1: User interface} communication refers to the different means to connect to the cluster on a particular node, including Graphical User Interfaces (GUIs) for cluster administration. 

\smallskip
\noindent
\emph{E.2.1.2.2: Resource management} communication can belong to two subclasses: base resource and agent. \emph{Base resource} describes the communication between the cluster resource management and those resources that are available as a standard (e.g., IP, CPU of a server). An \emph{agent} describes the communication between the cluster resource management and the agents that are responsible for managing application-specific resources (e.g., database, EA components) \cite{dake2008corosync,VeritasTechnologiesLLC2017}.

\smallskip
\noindent
\emph{E.2.1.3: Cluster} communication, also termed \emph{intra-cluster}, \emph{inter-node} (or resource group) communication, describes communication between cluster nodes. For a HAC, internal cluster communication is crucial. It is required for continuous communication between nodes regarding changes in configuration, the health status of nodes, quorum status, and failure notification. Furthermore, since cluster communication is often a basis for making necessary decisions by a cluster, the requirement for cluster communication is that it is enabled using an atomic (ordered) and reliable messaging scheme.  Even though several HAC solutions use different types of cluster communication, a strict definition can be used to distinguish the two main types: runtime and configuration. Thus, cluster communication deals primarily with the synchronisation of cluster \emph{configuration} and cluster \emph{runtime} data (e.g., the status of the nodes). 

\smallskip
\noindent
\emph{E.2.2: Method.} The types of communications utilise different transmission methods, and these methods can employ different protocols, such as UDP and TCP. Some HAC solutions employ custom protocols to meet the HAC-specific requirements, and an example is the Transparent Inter-Process Communication (TIPC) protocol,  used by OpenSAF \cite{toeroe2012service,maloy2004tipc}. The methods are further divided into four subclasses: multicast, broadcast, unicast, and IP socket.

\smallskip
\noindent
\emph{E.2.2.1: Multicast.} Multicast enables transmission from one node to multiple nodes. Thus, it can be characterised as a one-to-many (1:m) method. The receivers are usually a group of nodes, which means that a subset of cluster nodes can also be addressed \cite{forouzan2007data,dolev1996transis,marcus2003blueprints}.

\smallskip
\noindent
\emph{E.2.2.1.1: Atomic.} 
Atomic multicast (or total order multicast) implies that all nodes receive the same message in their sent order \cite{defago2004total}. 

\smallskip
\noindent
\emph{E.2.2.1.2: Virtual synchrony.} Virtual synchrony is an atomic multicast technology that supports reliable inter-process messaging. Corosync, the open-source communication protocol, employs the Totem Single-Ring Ordering and Membership (TOTEM) protocol, which is an example of implementation of virtual synchrony \cite{dake2008corosync}. Engelmann et al. \cite{engelmann2006symmetric} present a multi-node HAC solution for HPC that employs virtual synchrony to support state machine replication between the nodes in a symmetric active-active topology.

\smallskip
\noindent
\emph{E.2.2.2: Broadcast.} This method supports one-to-all (1:n) transmissions.  While multicast supports transmission to a group of nodes, broadcast transmits to all nodes \cite{forouzan2007data,schmidt2006high}.

\smallskip
\noindent
\emph{E.2.2.3: Unicast.} This method facilitates transmission between two nodes, and it is characterised as a one-to-one (1:1) transmission \cite{marcus2003blueprints,forouzan2007data}.

\smallskip
\noindent
\emph{E.2.2.4: IP socket.} An IP socket can also be used in some cases to facilitate communication between cluster nodes \cite{forouzan2007data,SIOSTechnologyCorp.2018}. However, the majority of the HACs do not support this method but rely on other methods.

\smallskip

Typically, cluster communication employs either multicast or broadcast, although unicast is also used in some cases. On the other hand, heartbeat communication employs either unicast or multicast  \cite{marcus2003blueprints}. The different communication types and methods may have limitations in cloud, fog and edge deployment environments. For example, routing a private IP address across subnets may not be possible in these environments, although such routing is required to failover from one subnet to another in a two-node cluster hosted in two data centres \cite{Infoscalecloud2020,SUSELLCCLOUD2021}. A solution could be to use an overlay network technique, which provides a network, e.g., an auxiliary network, on top of the main network, allowing routing across subnets~\cite{waldvogel2003efficient}, thus enabling seamless failovers.

\smallskip
\noindent
\textbf{E.3: Resource management.} Resources are structured hierarchically to form a resource group, and links between the resources define the relationships between the resources \cite{marcus2003blueprints}. 
 
\smallskip
\noindent
\emph{E.3.1: Type.} HACs can manage two types of resources: base resources and applications (Figure~\ref{fig:cluster-management}).

\smallskip
\noindent
\emph{E.3.1.1: Base resource.} A base resource is a standard building block (e.g., IP address, file system) \cite{critchley2014high,VeritasTechnologiesLLC2017,SUSELLC2017,OracleCorporation2017a}. A HAC can manage base resources without requiring additional tools. Hence, a distinction is made between base and application resources. While managing base resources is supported by all HAC solutions to different degrees, application support must be provided explicitly. 

\smallskip
\noindent
\emph{E.3.1.2: Application.} Application management is the capability to manage application-specific functionalities and features. Since each application must be handled individually, an extension to a HAC is usually required \cite{critchley2014high,VeritasTechnologiesLLC2013,IBMCorporation2017a}. Such addition is provided in the form of either an extension or an agent. 

\smallskip
\noindent
\emph{E.3.1.2.1: Agent-based.} Agents manage two main types of applications: application and database. \emph{Application} agents deal with managing several application-specific layers (e.g., application core of an ERP as in layer 2). \emph{Database} agents manage database-specific components (layer 3). The application agent functionality connects application-specific (this includes both types: database and application) configuration and procedures with the resource management module of a HAC and supports functionalities including
\cite{VeritasTechnologiesLLC2013,IBMCorporation2017a}:
\begin{itemize}
	\item Monitoring application-specific components
	\item An application-specific configuration, which can recognise the architecture of the application components
	\item Complying with application-specific dependencies
	\item Logging
	\item Procedures -- to stop and start related application components in a specific order
	\item Supporting Application Programming Interfaces (APIs) or specification by the application vendor
\end{itemize}

However, not all HAC solutions can support all applications, as each will require separate lifecycle management. When an application changes, for instance, when it is upgraded, the HAC application agent may also need to be updated to reflect the changes. Likewise, when the HAC solution is upgraded, the application agent may also need to be updated. Thus, supporting a large number of application agents could be connected to much effort. Furthermore, such support may be subject to licensing conditions, and HAC vendors could treat individual application support as an extension to license terms. 

\smallskip
\noindent
\emph{E.3.2: Method.} Two main methods are used when managing resources: policy- and rule-based.
\emph{Policy-based} resource management uses policies to configure conditions, and, when a particular condition is satisfied, appropriate action is triggered \cite{leangsuksun2004failure,VeritasTechnologiesLLC2017,IBMCorporation2017a}. On the other hand, \emph{rule-based} resource management uses one or more rules to make decisions and act upon them \cite{marcus2003blueprints}.

\subsection{F: Failure Detection and Recovery}
\begin{figure*}
	\centering
	\includegraphics[width=\linewidth]{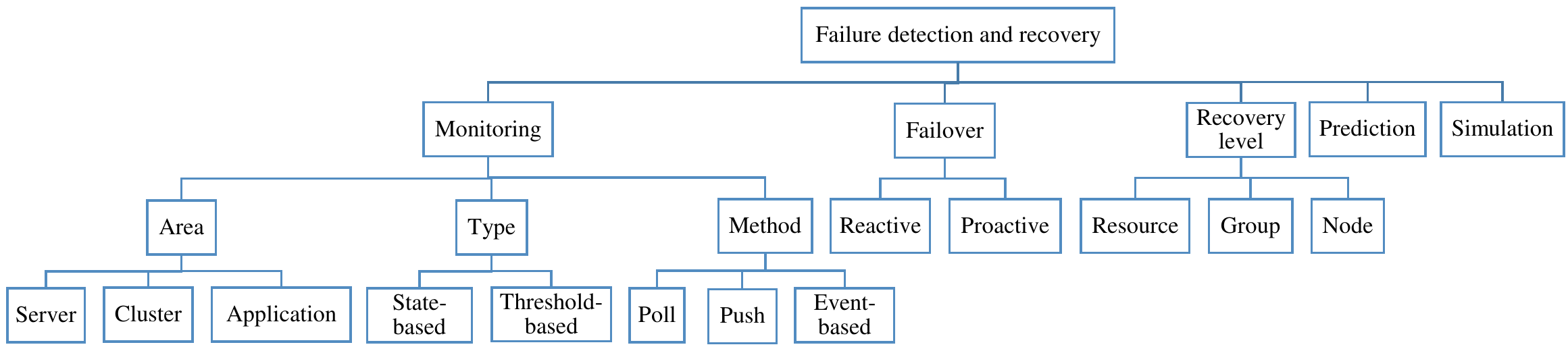}
	\caption{Failure detection and recovery.}
	\label{fig:failure-detection-and-recovery}
\end{figure*}

Failure detection implies detecting failures by monitoring and analysing monitoring output \cite{yang2020design}. If the monitoring identifies a status change in a resource or a resource group, it invokes recovery management to initiate a recovery.  If the recovery is not successful, the recovery manager may initiate a failover of a resource group or even a system; therefore, failover is part of recovery management. Figure~\ref{fig:failure-detection-and-recovery} depicts the top-level class with its subclasses.

\smallskip
\noindent
\textbf{F.1: Monitoring.} HAC failure detection and recovery monitoring can be further organised into subclasses depending on its area, type and method. 
The area describes the monitored domains, while the type of monitoring addresses monitoring from a configuration point of view. In most cases, HACs can provide support for specific monitoring metrics; however, if there is no support, a custom approach where HAC users define their own monitoring metrics is adopted. The monitoring scope may also vary and can range from the simple state monitoring of a resource to the monitoring of a resource in a detailed manner \cite{marcus2003blueprints,schmidt2006high,VeritasTechnologiesLLC2017,IBMCorporation2016a,FujitsuLimited2017}. Several research initiatives refer to the monitoring aspect of HACs as means to detect failures. Cheng et al. \cite{Cheng2005} present a state-based internal monitoring approach for the experimental cluster APCS+PEV, while Leangsuksun et al. \cite{leangsuksun2004failure} employ threshold-based monitoring for the cluster HA-OSCAR.

\smallskip
\noindent
\emph{F.1.1: Area.} The monitoring areas that a HAC can support play an important role in the overall solution. This is because monitoring is the process that collects details regarding monitored elements from different areas and delivers that data to the cluster management to make appropriate decisions. The areas that a solution can support can roughly be split into three subclasses: server, cluster, and application.

\smallskip
\noindent
\emph{F.1.1.1: Server.} Server-specific metrics focus on critical and non-critical monitoring elements of an operating system and a server level.  Examples of metrics are CPU utilisation and memory utilisation \cite{VeritasTechnologiesLLC2017,OracleCorporation2017a,IBMCorporation2016a,NECCorporation2017b}.

\smallskip
\noindent
\emph{F.1.1.2: Cluster.} Cluster monitoring implies that monitoring is enabled, even for the internal components of a HAC, including cluster-related processes and objects \cite{NECCorporation2017b,NECCorporation2017a,VeritasTechnologiesLLC2017}. This approach enables a HAC to distinguish between failures of cluster and application elements, thus preventing making incorrect decisions.

\smallskip
\noindent
\emph{F.1.1.3: Application.} Application monitoring is usually administered by an application-specific agent or an extension that is specifically designed to support a particular application and its architecture. This implies that an application agent is aware of the internals of the application \cite{OracleCorporation2016,IBMCorporation2017a}.

\smallskip
\noindent
\emph{F.1.2: Type.} A monitoring type describes how the state of the resources is measured. There are two types of monitoring: state- and threshold-based.

\smallskip
\noindent
\emph{F.1.2.1: State-based monitoring} uses the state of a resource as a monitoring metric, and the states can be as simple as “up” and “down,” or the monitoring can be more elaborate and contain more states \cite{VeritasTechnologiesLLC2017,SUSELLC2017,IBMCorporation2016a,IBMCorporation2017a}. 

\smallskip
\noindent
\emph{F.1.2.2: Threshold-based monitoring} uses a set of threshold values related to metrics \cite{ward2014observing}. As such, alerts with different severity levels can be generated, depending on which threshold is exceeded. While the threshold-based type gives the flexibility to configure monitoring at a granular level, it also adds complexity as the HAC must interpret all the different values and severity levels and act accordingly. One advantage is that a HAC will have more data that can be analysed, and decisions can be made at a granular level.

\smallskip
Even though state-based monitoring is the common type of monitoring, both monitoring types (and others) are sometimes combined. For example, OpenSAF HACs combine thre\-shold-based monitoring with a type called \emph{watermark} monitoring. The thre\-shold-based monitoring is used to monitor system resources, while the watermark monitoring is employed to register the highest and lowest utilisation per configured resource \cite{ServiceAvailabilityForum2011}.

\smallskip
\noindent
\emph{F.1.3: Method.} There are mainly three methods for monitoring, and they are push, poll, and event-based. \emph{Polling} implies that the monitoring module of HAC and agents poll for state changes of resources periodically \cite{ward2014observing,endo2016high}. On the other hand, \emph{push} implies monitoring data is pushed to the monitoring module or agents \cite{endo2016high, ward2014observing}. Such a setup will require additional enablers to interact with resources and push monitoring data to HAC agents. Polling is the most common method of monitoring \cite{endo2016high}, and it usually employs synchronous communication. However, this procedure is associated with a specific overhead. Therefore, other methods are studied by both industry and academia, and a technique that applies an \emph{event-based} design is viewed as less resource-demanding. One type of event-based monitoring employs an intermediate module that interfaces with an operating system to capture instantaneous notifications relevant, for instance, the state change of a process. It passes that to an appropriate module of a HAC. An example of such a setup is the Intelligent Monitoring Framework (IMF) by Veritas \cite{VeritasTechnologiesLLC2017}. The IMF has a monitoring feature integrated into an operating system for a particular resource so that state changes are captured instantaneously, and a relevant HAC agent is alerted. However, this approach requires specific development towards an operating system for a particular set of resources. Thus, IMF is not available for all types of resources but is being released gradually for different applications. Another recent development is to enable a HAC to interact with the monitoring feature of an operating system directly \cite{IBMCorporation2016a}, which means that the HAC needs only a slim variant of the monitoring module. The downside of this approach is that the HAC becomes highly dependent on the operating system and its developments.

\smallskip
\noindent
\textbf{F.2: Failover.} Failover management includes procedures for failover and failback, and all such actions are usually policy-driven. \emph{Policy-based} indicates that policies can be associated with events so that the appropriate policies are triggered whenever a related event occurs \cite{hiep2020dynamic}. Policies can be used, for example, to determine the target node for failover. Furthermore, policies can encode application-specific requirements, such as the order for starting up or shutting down resources. Failover management can further be split into two subclasses: reactive and proactive \cite{kaitovic2018impact}.

\smallskip
\noindent
\emph{F.2.1: Reactive.} A reactive measure uses policies to ensure the correct failover actions. There are two types of policies: static and dynamic. A static policy is created during the implementation or when applying manual changes, while a dynamic policy is created automatically by HACs to enforce policies based on runtime failure cases \cite{Beekhof2017}.

\smallskip
\noindent
\emph{F.2.2: Proactive.} A proactive measure assumes that a predictive model is employed to ensure that a failover can be initiated based on predictions \cite{kaitovic2018impact}. The predictions can, in turn, use policies to trigger the required actions \cite{leangsuksun2004failure}. However, the proactive approach could be a challenge in HAC environments that deal with complex EAs because all relevant layers must be addressed in such cases while evaluating the HAC behaviour. Therefore, all active HAC solutions employ only the reactive mechanism.

\smallskip
\noindent
\textbf{F.3: Recovery level.} The threefold strategy to manage failures is implemented using three recovery procedures: resource, resource group, and node (system) level \cite{schmidt2006high,VeritasTechnologiesLLC2017}. 

\smallskip
\noindent
\emph{F.3.1: Resource.} A resource-level recovery deals with recovery attempts on a resource level, implying re-initialisation of a failed resource while adhering to the dependency rules between resources. However, if this step fails, the failure is propagated to a resource-group level \cite{schmidt2006high}.

\smallskip
\noindent
\emph{F.3.2: Group.} A group-level recovery attempts to failover the entire resource group to a secondary node. However, if there are no available secondary nodes, an attempt to reinitialise the resource group within the same node can also be initiated. If a resource group have dependencies on other resource groups, it may lead to a node (system) recovery \cite{schmidt2006high}. 

\smallskip
\noindent
\emph{F.3.3: Node.} A node-level (system) recovery deals with failing over the resource groups to a secondary node. Moreover, a resource or resource group failure can also have a cascading effect due to dependencies, and, in such cases, it might lead to recovery on a node level. Since the previous node is labelled as "failed," policies may prevent any resources from being started there until that node is repaired \cite{schmidt2006high}.

\smallskip
\noindent

\textbf{F.4: Prediction.} Prediction in the context of HACs implies that prediction approaches are used to provide prediction at the different levels to improve the operations of HACs. One such example is to predict resource failures; however, current HACs do not commonly employ prediction. Some research initiatives explore the area of predicting failures, but often within a limited scope. An example of such an initiative is the HA-OSCAR project which assesses prediction by evaluating hardware component failures \cite{leangsuksun2004failure}. The research team used a Hardware Platform Interface (HPI), as specified by the Service Availability Forum, to identify hardware events and, subsequently, to analyse such data in order to provide predictions \cite{leangsuksun2004failure}. Similarly, Lee et al. \cite{lee2008stochastic} propose a stochastic prediction model for node failure or a node-switch interconnected system failure of HA-OSCAR head nodes. Leangsuksun et al. \cite{liu2003availability} have also explored a failure-repair model for predicting the availability of HA-OSCAR cluster by using Stochastic Reward Nets (SRNs). While many of the prediction models use HA-OSCAR as the platform, some initiatives explore other platforms. For example, Cheng et al. \cite{Cheng2005} have used a module for a custom cluster solution that detects sick nodes and subsequently uses a prediction method to forecast the time-to-failure of nodes. 

\smallskip
Both Veritas InfoScale Availability and Oracle Clusterware provide functionalities to simulate failures and observe potential paths to failovers \cite{VeritasTechnologiesLLC2017,OracleCorporation2017}. However, the objective of the subclass \textit{prediction} is to ensure that the wealth of information that HACs produce can be incorporated to predict failures or optimise failovers. An example of such an approach could be using prediction to optimise the quorum voting process by dynamically evaluating scenarios. 

\smallskip
\noindent
\textbf{F.5: Simulation.} A simulated cluster or cluster simulation is a feature to run cluster simulations to study the potential failover and recovery paths when failure is simulated at a resource, multiple resources, resource groups, or a system level. Simulating failures and studying the cluster behaviour is becoming increasingly important for the configuration and optimisation of complex HAC solutions. The feature can be provided: (1)~as part of HAC software; (2)~as a separate tool; and (3)~as part of the system management software.

Most HAC solutions provide a simulation feature with varying capabilities, and many simulators can be executed without interfering with the running cluster solutions. OpenSAF provides a simulated cluster (method 1) with the source code distribution. The resulting five-node simulated cluster can be brought up quickly to evaluate the cluster and perform functional and API tests. Further examples of simulated clusters include the simulator functionality provided by the ClusterLab stack (method 1), which can be used to simulate failures to study the potential failover paths. The simulator feature continues to add more features to enable HACs to deliver optimal services. Red Hat Enterprise Linux HA Add-On provides a command-based tool (method 1) to simulate recovery scenarios. Veritas provides a standalone simulator (method 2) for the Veritas cluster server (VCS) to simulate and test different failover situations and use the results to optimise the cluster configuration \cite{VeritasTechnologiesLLC2017}. IBM PowerHA SystemMirror for AIX delivers an advanced graphical cluster simulator as part of the IBM Systems Director, which is used to manage systems (method 3) \cite{quintero2015ibm}. The simulator supports creating and saving simulated cluster topologies. It enables conducting experiments with different configuration options, and one of the saved simulations can then be used to deploy the cluster when all the required components are in place. 

\subsection{G: Consistency and Integrity} 
\begin{figure*}
	\centering
	\includegraphics[width=\linewidth]{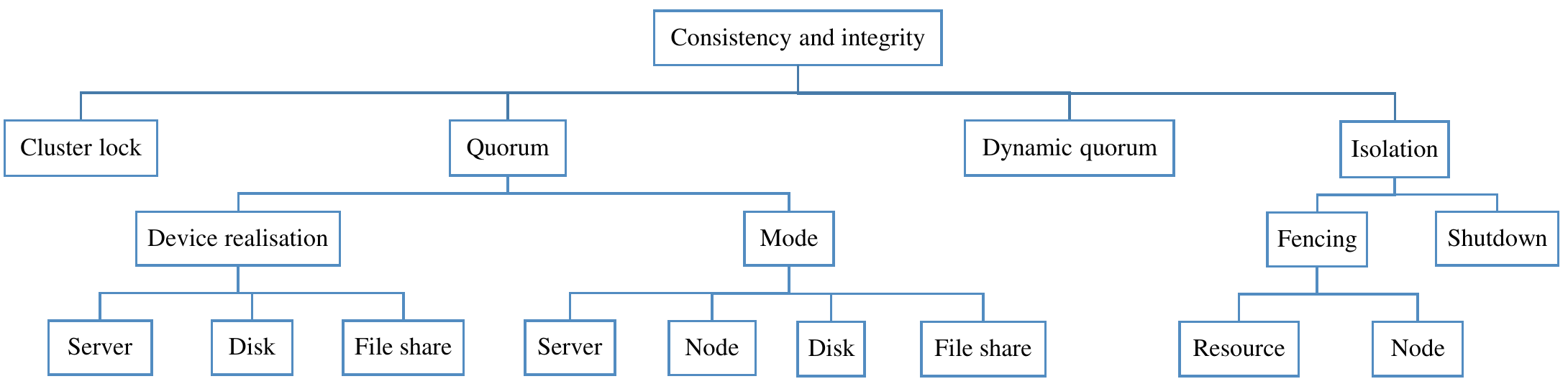}
	\caption{Consistency and integrity.}
	\label{fig:consistency-and-integrity}
\end{figure*}

Figure~\ref{fig:consistency-and-integrity} presents the top-level class \textit{consistency and integrity} and its subclasses. A HAC employs measures, such as a cluster lock or quorum, to preserve the data integrity of cluster resources and, most importantly, the clustered application by preventing harmful situations, for instance, \textit{split-brain} and \textit{amnesia}. 

\smallskip
\noindent
\textbf{G.1: Cluster lock.} Cluster lock is a technique used to lock cluster resources to a particular node, thus preventing other nodes from claiming the same resources. While a quorum-based approach could also be viewed as a cluster lock, a distinction is made to separate a quorum from a cluster lock. A cluster lock is a technique that does not employ a quorum-based approach but uses other means, such as a software-based lock mechanism. In the case of HACs, a distributed cluster lock is one such option, and an example is OpenSAF, which uses a global lock service to manage shared resources and ensure that only one node can access the resources at any given time \cite{toeroe2012service}. Such configurations are deemed \textit{quorum-less}.

\smallskip
\noindent

\textbf{G.2: Quorum.} A HAC quorum serves two purposes: 1)~maintaining cluster consistency by storing configuration and runtime data (e.g., cluster data)~\cite{gomes2021cloud}, and 2)~managing a voting system required in the event of a cluster partition.  For the latter purpose, the quorum hosts a voting mechanism in which every healthy and active node has a vote \cite{naor1998load,critchley2014high}. Furthermore, the quorum also has a vote, a potential decider, hence the alternative name \textit{tiebreaker}. Other names that are used to refer to the quorum mechanism are \textit{arbitrator}, \textit{witness} and \textit{voting system} \cite{schmidt2006high}. When a partition of a cluster occurs after the failure of one or more nodes, a quorum is gathered to decide which partition should have the quorum. To reach a quorum, a partition must have a majority of votes \cite{wang2015tradeoff}. The quorum service casting its vote can ensure that one of the partitions achieves this majority. Ultimately, the majority cluster is allowed to run the cluster. If a quorum cannot be reached, the surviving nodes will shut down to ensure cluster consistency. The quorum collaborates closely with the heartbeat mechanism, as the heartbeat is the method used to identify unhealthy nodes. Additionally, the quorum or a similar service is required for fencing, as the two often collaborate to determine a quorum and subsequent fencing. 

A quorum consists of a device and a process \cite{critchley2014high,vogels1998design,HewlettPackardEnterpriseDevelopmentL.P.2011}. A device describes where quorum elements are stored, and a device facilitates the process, which uses an algorithm to calculate votes to achieve a quorum. The process employs a mode to determine what policy to use when performing the quorum voting.

\smallskip
\noindent
\emph{G.2.1: Device realisation.} Three types of devices can be used by a HAC: server, disk, and file share. A quorum \emph{server} is a service that runs on a server that is usually hosted outside a cluster configuration \cite{critchley2014high}. The cluster is subsequently configured to connect to the quorum server. A \emph{disk-based quorum} is based on a disk, which can be either local or shared \cite{schmidt2006high,critchley2014high}. A \emph{file share} uses a shared file location, and it can be ideal for geographically distributed HACs since member nodes do not have access to a shared disk \cite{MicrosoftCorporation2011}. The prerequisite for all quorum devices is that they support concurrent access by all cluster members.

\smallskip
\noindent
\emph{G.2.2: Mode.} Four modes are possible:  server, node, disk and file share. The modes and the devices are an integral part of the quorum solution. However, the supported combination of devices and modes are specific to the different HAC solutions. The mode server uses the device \textit{server}, and the device \textit{Disk} is used by the mode Disk. Similarly, the device \textit{File share} is used by the mode File share. While the devices \textit{disk} and \textit{File share} imply that they are storage points that are managed by the quorum process, a \textit{quorum server} indicates an advanced device type. The mode node is implemented implicitly. Hence, it does not require any additional devices but uses the number of available nodes to decide, and the arrangement is referred to as the `majority node' mode.

\begin{table*}
	\sffamily
	\centering\footnotesize
	\caption{Quorum implementation with Windows server failover cluster (WSFC)}
	\label{tab:quorum-implementation-WSFC}%
\begin{tabular}{P{10em}P{8em}P{10em}P{12em}}
		\toprule
		Combination of Modes & Devices Realisation & Formula for Number of Nodes & Purpose \\
		\midrule
		Majority Node & Node only (implicit device) & $n = 2k + 1$ (odd numbers) & Survive failures of $(n -1)/2$ nodes. \medskip \\
		Node and Disk Majority & Node and disk & $n = 2k$ (even numbers) & Survive failures of $n/2$ nodes when disk is available. \medskip \\
		Node and File Share Majority & Node and file share & $n = 2k$ (even numbers) & Survive failures of $n/2$ nodes when file share is available. \medskip \\
		No Majority: Disk Only & Disk only & -     & Survive failures of $n-1$ nodes when disk is available. \\
		\bottomrule
	\end{tabular}%

\vspace*{-3mm}
\end{table*}%

A quorum can be set up in different ways, and, in some cases, the several modes of a quorum can be combined. For example, Windows Server Failover Clustering (WSFC) supports a combination of devices and modes, as detailed in Table~\ref{tab:quorum-implementation-WSFC} \cite{MicrosoftCorporation2011}. However, the same combination is not always supported by other HAC solutions, and an example of this is that the Serviceguard HAC does not recommend combining a quorum server with a quorum disk \cite{HewlettPackardEnterpriseDevelopmentL.P.2011}. There are new quorum device types introduced to meet the advances in IT. For example, Microsoft has introduced recently a new quorum device called \emph{cloud witness}, and the purpose is to support a server-based quorum in the Azure cloud, which could be ideal for cloud-based solutions \cite{MicrosoftCorporation2018}.

The standard for all explicit quorum devices is that they are placed outside a HAC to avoid creating a quorum device as a SPOF. Moreover, redundancy of quorum is also preferred because the quorum is a critical HAC functionality. For this reason, most current HAC solutions support a dynamic reconfiguration procedure for quorum devices, which enables adding or removing quorum devices without impacting the running clusters. While quorum is crucial for a two-node cluster, it can also be opted out using a different mechanism. Furthermore, when a cluster has more than two nodes, an explicit quorum device could become optional because that cluster can survive the failure of a single node. However, a configuration using the mode node is still required to achieve a quorum. The research in this area focuses on enabling probabilistic approaches. For instance, Malkhi et al.~\cite{malkhi2001probabilistic} have explored a probabilistic approach to address both benign server failures and arbitrary (Byzantine) ones.

\smallskip
\noindent
\textbf{G.3: Dynamic quorum.} While a quorum deals with static votes, a dynamic quorum calculates the number of votes and adjusts the quorum dynamically upon the failure of one or more nodes \cite{MicrosoftCorporation2018}. Thus, if a node is unavailable, it will effectively be out of the quorum voting process. This gives more flexibility to continue running a cluster even when other nodes fail. For example, dynamic quorum enables WSFC to run a cluster when only one node and a quorum device are available \cite{MicrosoftCorporation2016}.

\smallskip
\noindent
\textbf{G.4: Isolation.} HACs may ``isolate'' a particular node from the rest of the cluster, i.e., prevent it from allocating any resources. The objective of node isolation is to preserve data integrity by employing several mechanisms, such as putting a fence around a node (fencing) or shutting down a node.

\smallskip
\noindent
\emph{G.4.1: Fencing.}  There are two types of fencing, node-level and resource-level \cite{critchley2014high,VeritasTechnologiesLLC2017,SUSELLC2017,IBMCorporation2016a}. The common implementation is to employ the node-level fencing \cite{VeritasTechnologiesLLC2017,SUSELLC2017,IBMCorporation2016a}.

\smallskip
\noindent
\emph{G.4.1.1: Resource.} Resource-level fencing isolates one or more critical resources and, by doing so, renders a node unusable because the node cannot allocate resources. Resource-level fencing can be based on a SAN switch, allowing only one node to connect to the SAN-based storage or SCSI. SCSI-based fencing often uses a SCSI-3 option called persistent reservation, which means there can be only one SCSI-3 persistent reservation per disk at any given time, making it an efficient method for isolating disks \cite{VeritasTechnologiesLLC2017,SUSELLC2017,IBMCorporation2016a,zhu2020scsi3}. Since resource-level fencing is based on storage input/output (I/O), it is sometimes called \emph{I/O fencing} \cite{preslan2000scalability}.

\smallskip
\noindent
\emph{G.4.1.2: Node.} On the other hand, node-level fencing acts at a node-level and isolates or quarantines the node completely \cite{lumpp2008high}. In some cases, the node can be shut down instead, but the fencing functionality still manages the operation. Furthermore, the state of the fenced node is effectively changed so that it is no longer recognised as an active node by the cluster. Thus, the isolated node is not participating in any cluster operations.

\smallskip
\noindent
\emph{G.4.2: Shutdown.} A node shutdown is different from the shutdown procedure managed by the fencing functionality because it operates outside the fencing mechanism. This can be achieved by a HAC module that interacts with operating systems or servers using industry-standard specifications. Examples of APIs based on specifications are: Intelligent Platform Management Interface (IPMI) and vendor-specific embedded technology, such as Integrated Lights-Out (iLO) by HPE \cite{SUSELLC2017,ServiceAvailabilityForum2011,NECCorporation2017b,lee2021high}.

\subsection{H: Data Synchronisation.} 
Data synchronisation refers to the means, technologies and methods used to synchronise data between cluster nodes. The different layers of EAs require that data are synchronised to ensure consistency across all cluster nodes. Although a diverse range of synchronisation methods can be employed at the different layers, the overall responsibility for all layers managed by HACs lies with the HACs because they are responsible for failover management and ensuring data integrity. HACs may employ additional tools or features that come with the application components to facilitate data synchronisation. Hence, we identify three principal areas of data synchronisation:  

\begin{figure*}
	\centering
	\includegraphics[width=0.8\linewidth]{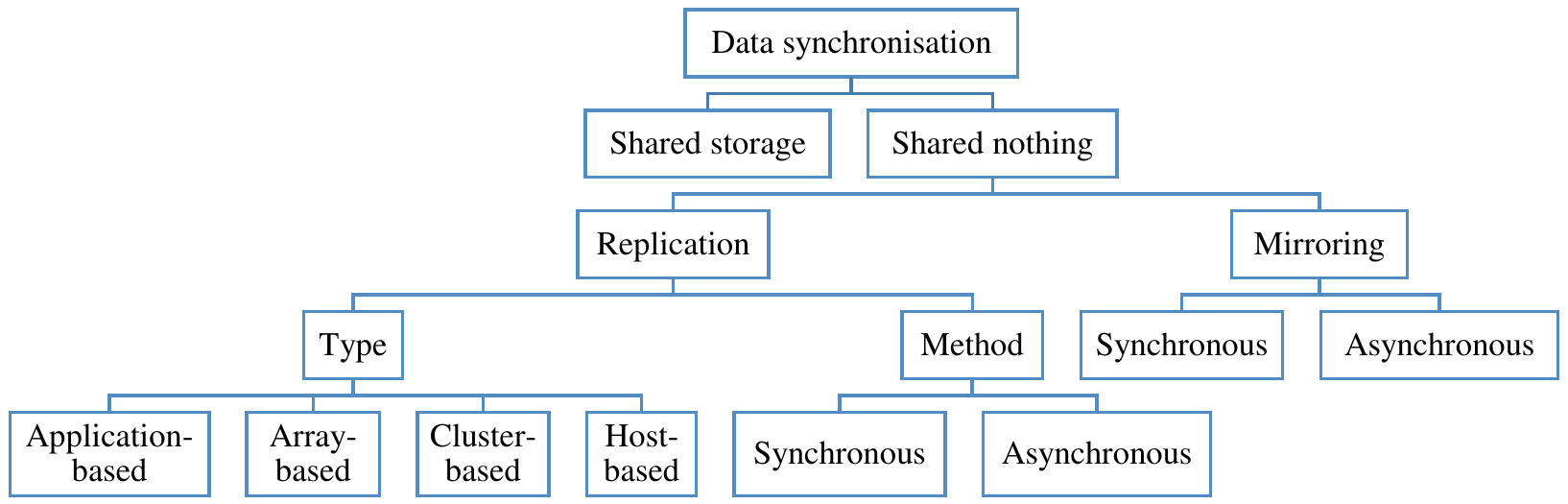}
	
	\vspace*{-2mm}
	\caption{Data synchronisation.}
	\label{fig:HAC-data-synchronisation}

	\vspace*{-2mm}
\end{figure*}

\renewcommand{\theenumi}{\arabic{enumi}}
\begin{enumerate}
\item \emph{Client-state} (i.e., session state replication) deals mainly with client connectivity (e.g., sessions), which means the client state of an application running on a primary node is synchronised with other cluster nodes \cite{rossi2005analyzing,van2020practical,mortazavi2020sessionstore}. Subsequently, a failover can occur seamlessly and without losing any connection data or affecting any active connections. Hence, other nodes can continue to support the connections instead.  Client-state synchronisation is widely employed in HAC for the application area \emph{network appliances} (e.g., firewalls) \cite{NOBLE2003191,CheckPointSoftwareTechnologiesLtd2018,PaloAltoNetworksInc.2018,fondo2020software}. 

\item \emph{Cluster-state} employs different methods to synchronise the state of a cluster, and it can be considered as an intra-cluster activity. For instance, OpenSAF uses a checkpoint service to record the state of an application or a service. Subsequently, states are replicated to a standby application or service that is hosted on the secondary node \cite{toeroe2012service}. A more advanced approach is a \emph{State Machine Replication} (SMR) which creates replicas of client and process states to one or more nodes deterministically \cite{le2016dynamic}, which can even support more comprehensive solutions such as databases \cite{pedone2003database,10.1145/3442197}. An example of SMR concerning a HAC is an implementation of a HAC for HPC, which employed SMR to synchronise states between nodes in a symmetric active-active topology \cite{engelmann2008symmetric}. A variation of SMR is \emph{Replicated State Machine} (RSM) employed by VCS to synchronise the resource status across all nodes \cite{VeritasTechnologiesLLC2017}.

\item \emph{Application-state}, on the other hand, implies that the data of an application that a HAC protects are synchronised to one or more nodes to support a possible failover. Hence, data synchronisation in this taxonomy refers implicitly to application-state synchronisation. Such synchronisation can occur at different levels, such as on an application or a file system level. 
\end{enumerate}

\begin{figure*}
	\centering
	\includegraphics[width=.55\linewidth]{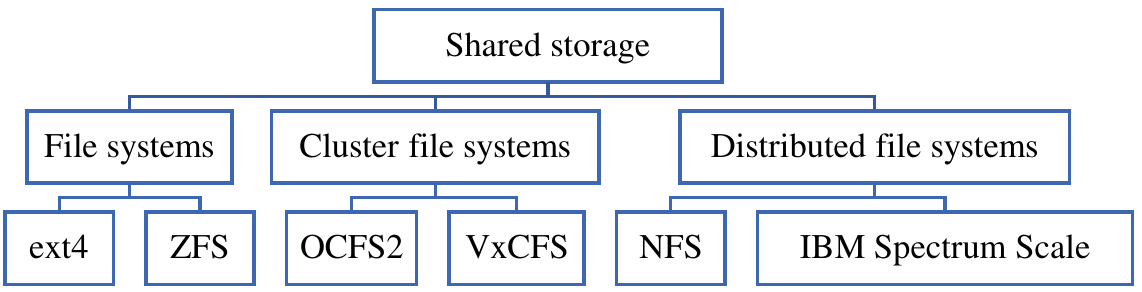}
	\caption{File systems related to shared storage. Key: ext4 -- Fourth extended file system, ZFS -- Z File System, OCFS2 -- Oracle Cluster File System, vxCFS -- Veritas Cluster File System, NFS -- Network File System, IBM Spectrum Scale -- A distributed file system, formerly called the General Parallel File System (GPFS)}
	\label{fig:shared-storage-file-systems}
\end{figure*}

The top-level class \textit{data synchronisation} is shown in Figure~\ref{fig:HAC-data-synchronisation}. It is further divided into two storage technologies, shared storage and shared-nothing. They both can be connected to a subclass of file systems which, in turn, can influence the configuration of a HAC. An example of file systems related to shared storage is presented in Figure~\ref{fig:shared-storage-file-systems}. Both cluster and distributed file systems support concurrent access and are ideal for sharing data between multiple nodes \cite{shi2020distributed}. A distributed file system can be deployed on the top of either shared storage or shared-nothing, and some file systems can be deployed on both. For example, IBM Spectrum Scale (formerly the General Parallel File System (GPFS)) file system can be deployed using both storage technologies \cite{IBMCorporation2017}.

\smallskip
\noindent
\textbf{H.1: Shared storage.} HAC solutions use several forms of shared storage. However, requirements for such implementation usually come from business requirements, such as supporting DR or geographically dispersed user groups. While shared storage might be ideal for the cluster types local and campus, metro and continental clusters require a different solution due to extended distances between nodes. Shared-nothing is an option in such cases. A hybrid approach is also possible, which means that shared storage and shared-nothing can support a combination of a local cluster (or campus) and continental cluster, as discussed in the topology section.
 
 \smallskip
\noindent
\textbf{H.2: Shared-nothing.} This setup assumes that there is no shared storage. Instead, each cluster node is connected to separate storage, which could either be SAN-based or based on local storage (e.g., Direct-attached storage (DAS)) \cite{marcus2003blueprints,schmidt2006high,critchley2014high}. However, EAs must explicitly support these kinds of setups. Moreover, there are also challenges with accessing shared storage in new and emerging technologies. For example, shared storage is limited in the public cloud, fog and edge computing environments \cite{rani2021storage}; hence, it becomes difficult to set up a HAC using shared storage. In such cases, replication between the individual storage units is required, and this has led to the new term  \textit{SANless} (SAN-Less) \cite{SIOSTechnologyCorp.2017a}. There are two techniques associated with the synchronisation of data in a shared-nothing setup: replication and mirroring.

\smallskip
\noindent
\emph{H.2.1: Replication.} Replication describes the process of replicating from a primary node to other nodes so that data is synchronised and consistent across all participating nodes \cite{marcus2003blueprints,schmidt2006high,critchley2014high,VeritasTechnologiesLLC2017}. Since there are different kinds of replications in HACs, we group them by type and method. The type describes the replication approaches, while the method specifies the execution technique.

\smallskip
\noindent
\emph{H.2.1.1: Type.} Four types of HAC replication are possible: application-based, array-based, cluster-based, and host-based.

\emph{Application-based replication} is set up at an application level, and it uses replication features that are provided natively by an application \cite{saxena2020cloud}. One of the nodes will be active in such a setup, while other nodes will be either warm standby or hot standby. To include an application in a HAC, explicit support for the application-specific replication feature by the HAC is required. Databases employ application-based replication to synchronise with standby databases \cite{Xiong2016,lee2017parallel,OracleCorporation2017a,lu2021epoch}. \emph{Array-based replication} is set up on a storage system level to enable synchronisation between two storage systems (e.g., SAN- or NAS-based) \cite{critchley2014high,saxena2020cloud}. Additional software may be required to facilitate array-based replication. 
In \emph{cluster-based replication}, the replication functionality is within a HAC and entirely administrated by the HAC \cite{NECCorporation2017b,SIOSTechnologyCorp.2017}. It means that the solution is independent of the operating environment or any other tool; instead, it relies on a high-speed network connection. \emph{Host-based replication} uses software tools on a host (server or nodes) to perform replication. An example is using a Linux logical volume manager (LVM) to set up replication between two logical volumes across two nodes \cite{critchley2014high,IBMCorporation2017b}. Tools that operate on an operating system level and are similar to host-based replication can also be included in this category \cite{zhu2020scsi3}. For instance, Gómez et al. \cite{gomez2014fault} use a software-based Distributed Replicated Block Device (DRBD) solution to enable replication between two volumes at a block-level in a virtual cluster setup.

\smallskip
\noindent
\emph{H.2.1.2: Method.} HACs can perform the replication synchro\-nously or asynchronously. 

\emph{Synchronous} replication waits until a write is completed and an acknowledgement is received from the other replication end, guaranteeing consistency between the two replication points \cite{critchley2014high}. From a transaction viewpoint, synchronous replication can support all the ACID properties. Therefore, no data loss is usually associated with it \cite{kanagavelu2013software,pohanka2020evaluation}. However, synchronous replication is challenging with extended distances. Nevertheless, modern techniques may offer solutions. For instance, Schmidt \cite{schmidt2006high} means that connections up to a distance of 100 km can be achieved by using dark fibre. This results in latencies of 0.5 $\mu$s, which is adequate for synchronous replication. On the other hand, \emph{asynchronous} replication does not wait until the writing is completed but gets an acknowledgement as soon as data is received at the second point \cite{critchley2014high,lu2021epoch}. As such, it may not comply with the ACID properties entirely, which, in turn, may result in data loss.

Different factors influence the selection of a method, and some of the critical factors are \cite{marcus2003blueprints,critchley2014high}: the distance between two nodes; the volume of data transported between nodes; type of data; frequency (continuous or burst); business requirements, such as DR. There is a network latency recommendation for synchronous replication, as it implies real-time mirroring, while asynchronous replication does not have the same kind of rigorous requirement \cite{critchley2014high,pohanka2020evaluation}.

\smallskip
\noindent
\emph{H.2.2: Mirroring.} In some cases, the terms replication and mirroring are used interchangeably.  For example, a host-based mirroring of a file system can also be referred to as file system replication. However, in other cases, a few differences can be observed; for instance, mirroring may differ by not having a running instance on the standby node \cite{critchley2014high,schmidt2006high}. Mirroring can be performed synchronously or asynchronously~\cite{marcus2003blueprints,critchley2014high,schmidt2006high}. 

\emph{Synchronous} mirroring ensures that the mirroring process waits until a write is completed and committed on the standby node and an acknowledgement is sent back. This method secures consistency of data between two nodes. An \emph{asynchronous} mirroring process, on the other hand, does not wait until a write is ended on the secondary node. This approach may result in data loss when the primary node fails abruptly.

\section{Survey of High-availability Clusters \label{sec:survey}}

\subsection{Selection of HACs for the Survey}
\label{sec:approach-selecting-surveyed-HACs}
\begin{table*}
\caption{Selection questions}
\label{tab:HAC-survey-selection-questions}
	\centering\footnotesize
\begin{tabular}{P{.2cm}P{5.5cm}P{6.5cm}}
    \toprule
        \textbf{No} & \textbf{Question} & \textbf{Evaluation Parameters}\\
    \midrule
        Q1 & Support for enterprise class databases? & SAP ASE, DB2, HANA, Informix, MySQL, Oracle, PostgreSQL, SQL Server, Teradata \\
        Q2 & Support for EAs? & Oracle Siebel Customer relationship management (CRM), Oracle†, SAP† , Others†, WebSphere  \\
        Q3 & Multi-tier support for EAs? & X-Yes, N-no, ?-no information \\
        Q4 & Enterprise support provided? & 24x7x365 \\
        Q5 & Application features can be supported by further developments? & X-Yes, N-no, ?-no information \\
        Q6 & Support for disaster recovery? & X-Yes, N-no, ?-no information \\
        Q7 & Support for virtualization? & X-Yes, N-no, ?-no information \\
        Q8 & Cloud support? & X-Yes, N-no, ?-no information \\
        Q9 & Support for enterprise operating environments? & AIX, HP-UX, IBM i, Linux, Solaris, Windows \\
        Q10 & Support for multiple platforms? & Power, SPARC, x86 \\
        Q11 & Support for large-scale clusters? & Number of nodes \\
        Q12 & Support for multiple topologies? & Active-active, application-based, server-based, N-to-N, active-passive, N+1, N+M, N-to-1 \\
        Q13 & Support for availability level? & Minimum 99.9\% \\
        Q14 & Active lifecycle management? & X-Yes, N-no, ?-no information 
        \\
    \bottomrule
                  \multicolumn{3}{l}{†  - any of the business suite EAs (e.g., ERP)}
    \end{tabular}
  \end{table*}

\begin{table*}[]
\centering\footnotesize
\caption{Eliminated HAC solutions in the six-step approach for selecting HACs for survey}
\label{tab:eliminated-HAC-solutions-survey}
\begin{tabular}{@{}ll@{}}
\toprule
Product        & Reason(s) for Elimination \\ \midrule
Apache Mesos \cite{delvalle2017electron}   & focus on specific IT solutions (HPC) \\
DxEnterprise \cite{DxEnterprise}   & lack of EA support; insufficient information available to evaluate the solution properly  \\
everRun \cite{everRun}       &  lack of EA support; insufficient information available to evaluate the solution properly \\
HA-OSCAR \cite{haddad2003ha}      &  no longer active \\
Kimberlite \cite{leangsuksun2004failure}     & no longer active \\
Linux FailSafe \cite{leangsuksun2004failure}  & no longer active \\ \bottomrule
\end{tabular}
\end{table*}

We selected the relevant HACs for our survey using a systematic approach that comprised the following six steps.

\emph{Step~1. Identification of HACs that support enterprise applications (EAs).} 
Our survey focused on HACs that can protect EAs. However, only a limited number of HACs support EAs due to the complex composition of EAs, which are multi-tiered and multi-layered. We identified likely candidate HACs using comprehensive research reports \cite{IDCCorporation2016,prior2001enterprise} and articles \cite{leangsuksun2004failure,engelmann08symmetric3,rabbat2001high,corsava2003intelligent}, resulting in 23 candidate HAC solutions.

\smallskip
\noindent 
\emph{Step~2. Identification of relevant EAs and databases.} 
In this step, we gathered information for assessing the applicability of each HAC solution to distinct layers of enterprise applications. To this end, we used relevant research and analysis reports, e.g.~\cite{GartnerInc.2017,GartnerInc.2017erp,GartnerInc.2016crm}, to identify the databases and EAs listed next to questions Q1 and Q2 from Table~\ref{tab:HAC-survey-selection-questions}.

\smallskip
\noindent 
\emph{Step~3. Elimination of HACs not supported by EA vendors}.
In this step, we used the lists of supported HACs released regularly by enterprise application vendors, e.g.~\cite{SAPSE2018,IBMHAC2018}, to check which HAC solutions are supported (and sometimes certified) by these EA vendors. ``Supported” HAC solutions are solutions that fulfil the requirements of the application vendor for a specific application, with the added implication that support channels have been established between the vendors.

We assessed the candidate HAC solutions using the following criteria to narrow down the list:
\begin{enumerate}
\item Does the HAC solution being assessed focus on only specific IT solutions (such as HPC or Hadoop)?
\item Is the HAC solution no longer active, implying that the product lifecycle has ended or the research project that developed it has ended?
\item Is the information available to analyse the HAC solution properly insufficient?
\item Are EAs supported by the HAC solution, or is it the case that the information available cannot be used to conclude whether EAs are supported or not?
\end{enumerate}
We eliminated all the candidate solutions for which one or several of these questions were answered affirmatively. As a result, six HAC candidates were removed in this step, and we proceeded with the remaining 17 candidates. We made an exception for two of the candidate HAC solutions for the reasons described below: 

\begin{itemize}
 \item OpenSAF does not provide enterprise support directly. Nevertheless, we retained OpenSAF because of its stability as a HAC \cite{kanso2013achieving,khan2017comparing}. Besides, application support can be developed individually with OpenSAF, meaning that an OpenSAF HAC can be used to support enterprise-class databases and applications. 
 \item Similarly, the ClusterLabs stack does not suport enterprise applications on its own. However, we retained the ClusterLabs stack because it provides the core components for two other selected solutions, SUSE Linux Enterprise High Availability Extension and Red Hat High Availability Add-On. This implies that customisation and further developments are possible using it.
 \end{itemize}

 \smallskip
\noindent 
\emph{Step~4. Retention of only HACs that support automatic failover}.
We used this filter to retain only the HAC solutions that support automatic failover, which is crucial for an EA to minimise downtime. All 17 candidates support automatic failover; hence, all were retained.

\smallskip
\noindent 
\emph{Step~5. Design of additional questions for the selection and evaluation of HACs}. In this step, we created the questions to evaluate the HACs. The queries reflected the typical requirements of EAs \cite{franke2011optimal}, and the objective was to select those HACs that could respond to most of the questions positively. The set of questions is listed in Table~\ref{tab:HAC-survey-selection-questions}. 

\smallskip
\noindent 
\emph{Step~6. Selection of the set of HAC solutions for the survey}. We selected all the HAC solutions that can support EAs and fulfil the additional criteria from the questions Q1--Q14 in Table~\ref{tab:HAC-survey-selection-questions}, where a positive response for any of the ``evaluation parameters'' from questions Q1, Q2, Q9, Q10 and Q12 was deemed sufficient to consider that a HAC solution met the criterion associated with that question. The result of the HAC selection is presented in Table~\ref{tab:HAC-evaluation}, which comprises 17~HAC solutions for which we obtained positive responses to all queries and products, while noting the following exception:

 \begin{itemize}
 \item DR support (question~Q6) for the following solution was unclear or not available:  ApplicationHA, Clusterware 12c, Primecluster, RSF-1, SafeKit and WSFC. 
 \end{itemize}

For completeness, we also provide, in Table \ref{tab:eliminated-HAC-solutions-survey}, a list of the six HAC solutions considered initially but eliminated in Step~3 of our selection approach. For each of these solutions, Table \ref{tab:eliminated-HAC-solutions-survey} also provides a summary of the reasons for its elimination from the survey.

\subsection{HAC Analysis Methodology}
\label{sec:HAC-analysis-methodology}
We used a hybrid methodology for the analysis of the 17~HAC solutions from Table~\ref{tab:HAC-evaluation}. As a first step, we created a comprehensive spreadsheet and an online questionnaire covering our entire HAC taxonomy, which we used as a basis for the survey. In the second step, we populated the spreadsheet entries for 17 HACs by analysing product documentation, technical white papers, case studies, books, and articles as a primary analysis method. We noticed several inconsistencies between the different materials for the same edition and version of a HAC solution. To resolve these inconsistencies, we crosschecked the results by using a diverse set of materials (e.g., reference guides, technical manuals and documentation) whenever inconsistencies were observed. Finally, as a secondary method, we prepared an email that described what we were trying to achieve. We sent it to all the vendors of selected HACs, particularly to the experts responsible for the HAC products. After two weeks, we sent a reminder to those who did not reply to our original invitation; a second reminder was sent after an additional two weeks. After six weeks, we collected the data provided by the vendors and transferred it to a spreadsheet.

\twocolumn[
  \begin{@twocolumnfalse}
{\footnotesize
\begin{longtable}[c]{|P{.152\textwidth}*{4}{|P{.01\textwidth}}|P{.01\textwidth}*{6}{|P{.01\textwidth}}|P{.01\textwidth}*{5}{|P{.01\textwidth}}|}
		\caption{Evaluation of selected HAC solutions (X -- Yes; N -- No;	
    	? -- No information)}
		\label{tab:HAC-evaluation}\\
		\hline
		 {Question No} & \rotatebox{90}{ApplicationHA} & \rotatebox{90}{Clusterware} & \rotatebox{90}{EXPRESSCLUSTER X} & \rotatebox{90}{InfoScale Availability} & \rotatebox{90}{OpenSAF} & \rotatebox{90}{ClusterLabs stack} & \rotatebox{90}{PowerHA SystemMirror} & \rotatebox{90}{PRIMECLUSTER} & \rotatebox{90}{\parbox[c]{6cm}{\vfill Red Hat High Availability Add-On}} & \rotatebox{90}{RSF-1}  & \rotatebox{90}{SafeKit} & \rotatebox{90}{Serviceguard} & \rotatebox{90}{SIOS Protection Suite} & \rotatebox{90}{Solaris Cluster} & \rotatebox{90}{\parbox[c]{6.5cm}{SUSE Linux Enterprise High Availability Extension}} & \rotatebox{90}{Tivoli System Automation for Multiplatforms (SA MP)} & \rotatebox{90}{\parbox[c]{6.5cm}{Windows Server Failover Clustering (WSFC)}} \\ \hline
		\endfirsthead
		Q1  & X & X & X & X & X & X & X & X & X & X & X & X & X & X & X & X & X \\ \hline
		Q2  & X & X & X & X & X & X & X & X & X & X & X & X & X & X & X & X & X \\\hline
		Q3  & X & X & X & X & X & X & X & X & X & X & X & X & X & X & X & X & X \\\hline
		Q4  & X & X & X & X & N & N & X & X & X & X & X & X & X & X & X & X & X \\\hline
		Q5  & X & X & X & X & X & X & X & X & X & X & X & X & X & X & X & X & X \\\hline
		Q6  & ? & ? & X & X & X & X & X & ? & X & ? & ? & X & X & X & X & X & ? \\\hline
		Q7  & X & X & X & X & X & X & X & X & X & X & X & X & X & X & X & X & X \\\hline
		Q8  & X & X & X & X & X & X & X & X & X & X & X & X & X & X & X & X & X \\\hline
		Q9  & X & X & X & X & X & X & X & X & X & X & X & X & X & X & X & X & X \\\hline
		Q10  & X & X & X & X & X & X & X & X & X & X & X & X & X & X & X & X & X \\\hline
		Q11  & X & X & X & X & X & X & X & X & X & X & X & X & X & X & X & X & X \\\hline
		Q12  & X & X & X & X & X & X & X & X & X & X & X & X & X & X & X & X & X \\\hline
		Q13  & X & X & X & X & X & X & X & X & X & X & X & X & X & X & X & X & X \\\hline
		Q14  & X & X & X & X & X & X & X & X & X & X & X & X & X & X & X & X & X \\\hline
		\end{longtable}
					}%
  \end{@twocolumnfalse}
  ]
Despite assurance from multiple vendors, we managed to obtain a response from only one vendor, High Availability for the HAC RSF-1. Subsequently, we transferred all collected data to the spreadsheet for conducting further analysis.

\subsection{HAC Survey Results}
\label{sec:Survey-Results}
We used the taxonomy to establish the characteristics of the 17 end-to-end HAC solutions selected for the survey. The outcome of the survey is presented in Table~\ref{tab:outcome-survey}, starting with general information about each HAC solution (i.e., version and vendor) in the second and third row. The remaining rows from the table present the main results of the survey, organised in the same way as our HAC taxonomy. The results are analysed in Section~\ref{sec:analysis-survey-results}. 

The surveyed HAC solutions usually consist of multiple editions with varying features, some of which are subject to additional licensing. Our survey covers only advanced editions that include most of the features. As even advanced editions do not support all the features when different operating systems and platforms are considered, we provide details about the limitations relating to the individual HACs where applicable (as footnotes at the end of Table~\ref{tab:outcome-survey}). 

As discussed, a HAC vendor may enforce further constraints by stating explicitly what version and edition of an EA are supported. Likewise, an EA vendor may list what HACs are supported by a particular EA version and edition. Many combinations of EA version, database version, HAC version and edition, operating system version, and platform make it challenging to crosscheck every single combination. Therefore, only the relevant EAs and databases are included in Table~\ref{tab:outcome-survey}. 

\subsection{Analysis of the Survey Results}
\label{sec:analysis-survey-results}

\begin{figure*}
	\centering
	\includegraphics[width=\linewidth]{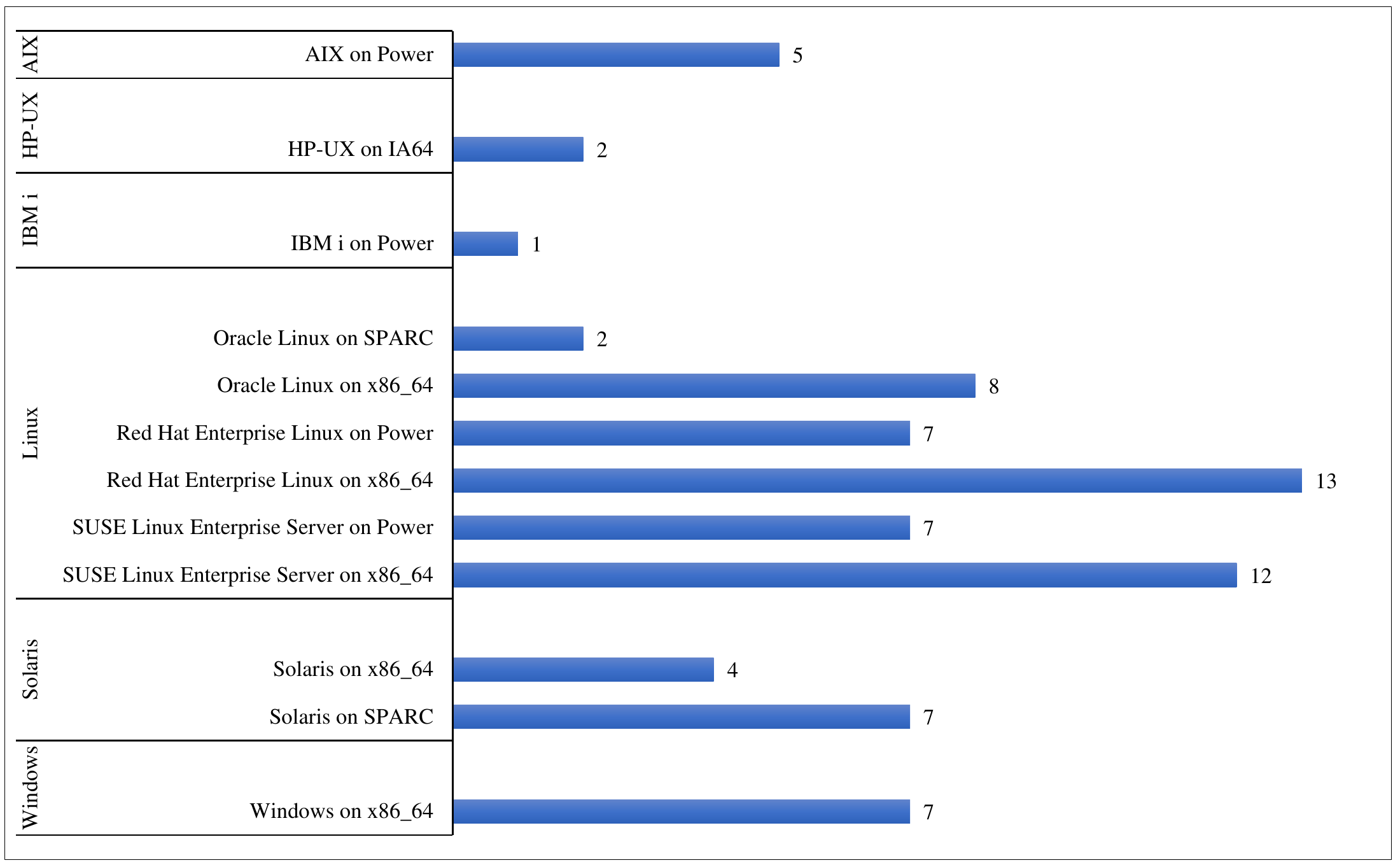}
	\caption{Platform and operating system support of the surveyed high availability clusters (HACs) grouped by operating system.}
	\label{fig:distribution-operating-system-platform-survey}
\end{figure*}

The distribution of the operating system and platform support for the surveyed HACs is shown in Figure~\ref{fig:distribution-operating-system-platform-survey} grouped by the operating system. Linux is the dominating operating system, and 15 solutions support Linux, out of which 12 support SUSE Linux on an x86-based platform, seven support SUSE Linux on Power-based platforms. Similarly, Red Hat Linux supports 13 HACs on the x86 platform and seven on the Power platforms. Oracle Linux is supported by 8 HAC solutions on x86 platforms, while only two support it on the SPARC platform. Solaris operating system is supported by seven HACs on the SPARC platform, while only four support Solaris on the x86 platform. Seven solutions support windows, and the platform is always x86. Five HACs support AIX on power, and only two HACs support HP-UX on the IA64 platform. Lastly, the rare environment is the IBM i  operating system on the Power platform, which is only supported by one HAC solution.

The surveyed HAC solutions can be divided into two groups. The first group, comprising 14 of the 17 surveyed solutions, comprises the HACs marked with a star `*' in Table~\ref{tab:surveyed HACs-versions-vendors}.
\twocolumn[
  \begin{@twocolumnfalse}
{
\scriptsize
	\begin{longtable}[c]{|P{.152\textwidth}*{4}{|P{.01\textwidth}}|P{.01\textwidth}*{6}{|P{.01\textwidth}}|P{.01\textwidth}*{5}{|P{.01\textwidth}}|}
		\caption{Outcome of the survey}
		\label{tab:outcome-survey}\\
		\hline
		 {Taxonomy} & \rotatebox{90}{ApplicationHA $^1$} & \rotatebox{90}{Clusterware} & \rotatebox{90}{EXPRESSCLUSTER X} & \rotatebox{90}{InfoScale Availability} & \rotatebox{90}{OpenSAF} & \rotatebox{90}{ClusterLabs stack} & \rotatebox{90}{PowerHA SystemMirror $^1$} & \rotatebox{90}{PRIMECLUSTER} & \rotatebox{90}{\parbox[c]{4cm}{\vfill Red Hat High Availability Add-On}} & \rotatebox{90}{RSF-1 $^{23}$} & \rotatebox{90}{SafeKit} & \rotatebox{90}{Serviceguard} & \rotatebox{90}{SIOS Protection Suite} & \rotatebox{90}{Solaris Cluster} & \rotatebox{90}{\parbox[c]{4.8cm}{SUSE Linux Enterprise High Availability Extension}} & \rotatebox{90}{Tivoli System Automation for Multiplatforms (SA MP)} & \rotatebox{90}{\parbox[c]{4.8cm}{Windows Server Failover Clustering (WSFC)}} \\ \hline
		\endfirsthead
		\multicolumn{18}{c}%
		{\tablename\ \thetable\ -- \textit{Continued from previous page}} \\[1ex]
		\endhead \multicolumn{18}{r}{\textit{Continued on next page}} \\
	    \endfoot
	    \endlastfoot
		{Version}  & \rotatebox{90}{6.2} & \rotatebox{90}{12c} & \rotatebox{90}{3.3} & \rotatebox{90}{7.3.1} & \rotatebox{90}{5.17.07} & \rotatebox{90}{2.3.2} & \rotatebox{90}{7.2.1} & \rotatebox{90}{4.5} & \rotatebox{90}{7.0} & \rotatebox{90}{3.9.10} & \rotatebox{90}{7.2} & \rotatebox{90}{A.12.20$^{24}$} & \rotatebox{90}{9.2} & \rotatebox{90}{4} & \rotatebox{90}{12} & \rotatebox{90}{4.1} & \rotatebox{90}{2016} \\ \hline
		{Vendor}  & \rotatebox{90}{Veritas} & \rotatebox{90}{Oracle} & \rotatebox{90}{NEC} & \rotatebox{90}{Veritas} & \rotatebox{90}{SA Forum} & \rotatebox{90}{ClusterLabs} & \rotatebox{90}{IBM} & \rotatebox{90}{Fujitsu} & \rotatebox{90}{Red hat} & \rotatebox{90}{High-Availability} & \rotatebox{90}{Evidian} & \rotatebox{90}{HPE} & \rotatebox{90}{SIOS} & \rotatebox{90}{Oracle} & \rotatebox{90}{SuSE} & \rotatebox{90}{IBM} & \rotatebox{90}{Microsoft} \\ \hline
\rowcolor{myblue}

		A: Deployment Patterns$^{27}$  &  &  &  &  &  &  &  &  &  &  &  &  &  &  &  &  &  \\ \hline
\rowcolor{myblue}		
		OS and platform   &  &  &  &  &  &  &  &  &  &  &  &  &  &  &  &  &  \\ \hline
		AIX on Power   & X $^1$ & X$^4$ & NS & X & NS & NS & X & NS & NS & NS & NS$^{22}$ & NS & NS & NS & NS & X & NS \\ \hline
		HP-UX on IA64  & NS & X & NS & NS & NS & NS & NS & NS & NS & NS & NS & X & NS & NS & NS & NS & NS \\ \hline
		IBM i on Power  & NS & NS & NS & NS & NS & NS & X & NS & NS & NS & NS & NS & NS & NS & NS & NS & NS \\ \hline
		Oracle Linux on SPARC  & NS & NS & NS & NS & X?$^8$ & X & NS & NS & NS & NS & NS & NS & NS & NS & NS & NS & NS \\ \hline
		Oracle Linux on x86\_64  & X $^1$ & X$^4$ & X & X & X & X & NS & NS & NS & X & NS & NS & X & NS & NS & NS & NS \\ \hline
		Red Hat Enterprise Linux on Power  & NS & NS & X & NS & X?$^8$ & X & X & NS & X & X & NS & NS & NS & NS & NS & X & NS \\ \hline
		Red Hat Enterprise Linux on x86\_64  & X $^1$ & X & X & X & X & X & NS & X & X & X & X & X & X & NS & NS & X$^{10}$ & NS \\ \hline
		Solaris on SPARC  & X $^1$ & X & NS & X & NS & NS & NS & X & NS & X & NS & NS & NS & X & NS & X & NS \\ \hline
		Solaris on x86\_64  & NS & X & NS & X & NS & NS & NS & NS & NS & X & NS & NS & NS & X$^{11}$ & NS & NS & NS \\ \hline
		SUSE Linux Enterprise Server on Power  & NS & NS & X & NS & X?$^8$ & X & X & NS & NS & X & NS & NS & NS & NS & X & X & NS \\ \hline
		SUSE Linux Enterprise Server on x86\_64  & X $^1$ & X & X & X & X & X & NS & X & NS & X & NS$^{22}$ & X & X & NS & X & X$^{10}$ & NS \\ \hline
		Windows  & X $^1$ & X & X & X & NS & NS & NS & NS & NS & NS & X & NS & X & NS & NS & NS & X \\ \hline
		Support for virtualized environments  & X & X & X & X & X & X & X & X & X & X & X & X & X & X & X & X & X \\ \hline
		Supported virtual solutions (E-Xen, H-Hyper-V, K-KVM, O-Others, V-VMware) & E, H, K, O, V & O & E, H, V & H, K, O, V & K, O, V, X & E, H, K, V & O$^9$ & K, O, V & K & ? & H & H, K, V & E, H, K, V & O & E, K & K, O, V & H, V \\ \hline
		Maximum number of nodes per cluster   & ? & 64 $^5$ & 32 & 128 & 100 & 32 & 16 & 16 & 16 & 64 & ? & 16/32 $^{25}$ & 32 & 8/16 $^{26}$ & 32 & 32/ 130$^{14}$ & 64 \\ \hline
		
		B: Application Areas (EA category, B- Business-critical, T-telecom (carrier-grade)) & B & B & B & B & T, B & B & B & B & B & B & B & B & B & B & B & B & B \\ \hline
		\rowcolor{myblue}
		C: Type of cluster  &  &  &  &  &  &  &  &  &  &  &  &  &  &  &  &  &  \\ \hline
		C.1: Local  & X & X & X & X & X & X & X & X & X & X & X & X & X & X & X & X & X \\ \hline
		C.2: Campus  & X & X & X & X & X & X & X & X & X & X & X & X & X & X & X & X & X \\ \hline
		C.3: Metro  & ? & X & X & X & x & X & X & NS? & X & X? & NS? & X & X & X & X & x & X \\ \hline
		C.4: Continental   & ? & NS? & X & X & x & X & X & NS? & X & NS & NS? & X & X? & X & X & x & NS? \\ \hline
	\rowcolor{myblue}
		D: Topology   &  &  &  &  &  &  &  &  &  &  &  &  &  &  &  &  &  \\ \hline
	\rowcolor{myblue}	
		D.1: Symmetric   &  &  &  &  &  &  &  &  &  &  &  &  &  &  &  &  &  \\ \hline
	\rowcolor{myblue}
		D.1.1: Active-active  &  &  &  &  &  &  &  &  &  &  &  &  &  &  &  &  &  \\ \hline
		D.1.1.1: Application-based   & NS & X & X$^7$ & X$^7$ & X$^7$ & X$^7$ & X$^7$ & X$^7$ & X$^7$ & X$^7$ & X$^7$ & X$^7$ & X$^7$ & X$^7$ & X$^7$ & ? & X$^7$ \\ \hline
		D.1.1.2: Server-based   & X & X & X & X & X? & X & X & X & X & X & X & X & X & X & X & X & X \\ \hline
		D.1.2: N-to-N   & X? & ? & X & X & X & X & X? & X? & ? & NS & ? & X? & X & X & X & ? & X \\ \hline
		\rowcolor{myblue}
		D.2: Asymmetric   &  &  &  &  &  &  &  &  &  &  &  &  &  &  &  &  &  \\ \hline
		D.2.1: Active-passive   & X & X & X & X & X & X & X & X & X & X & X & X & X & X & X & X & X \\ \hline
		D.2.2: N-to-1   & X? & ? & X & X & X & X & X & ? & X & X & X & X & X & X & X & ? & NS \\ \hline
		D.2.3: N+1  & ? & ? & X & X & X & X & X? & ? & ? & X & ? & X? & X? & X & X & ? & NS? \\ \hline

	\end{longtable}
    }%
      \end{@twocolumnfalse}
  ]

\twocolumn[
  \begin{@twocolumnfalse}
{
\scriptsize
	\begin{longtable*}[c]{|P{.152\textwidth}*{4}{|P{.01\textwidth}}|P{.01\textwidth}*{6}{|P{.01\textwidth}}|P{.01\textwidth}*{5}{|P{.01\textwidth}}|}
	\hline
			D.2.4: N+M   & ? & ? & X & X & X & X & X? & ? & ? & X & X & X? & X? & X & X & ? & NS? \\ \hline
\rowcolor{myblue}
		E: Cluster management   &  &  &  &  &  &  &  &  &  &  &  &  &  &  &  &  &  \\ \hline
		\rowcolor{myblue}
		E.1: Cluster data  &  &  &  &  &  &  &  &  &  &  &  &  &  &  &  &  &  \\ \hline
		E.1.1: Configuration (D-Disk or file share, F-File, M-memory) & D,F & F & F & F & F & F &D,F & F & F & D,F & F & F & D,F & F & F & F & F \\ \hline
		E.1.2:  Runtime (D-Disk or file share, F-File, M-memory) & D, M? & F, M? & F, M? & F, M? & F, M? & F, M & D,F,M? & F, M? & F, M? & D, M? & F, M? & F, M? & D,M? & F, M? & F, M? & F, M? & F, M? \\ \hline
		\rowcolor{myblue}
		E.2: Communication  &  &  &  &  &  &  &  &  &  &  &  &  &  &  &  &  &  \\ \hline
		\rowcolor{myblue}
		E.2.1: Type &  &  &  &  &  &  &  &  &  &  &  &  &  &  &  &  &  \\ \hline
	\rowcolor{myblue}
E.2.1.1: Heartbeat  & X$^{16}$ & X & X & X & NS & X & X & X & X & X & X & X & X & X & X & X & X \\ \hline
E.2.1.1.1: LAN-based  & X & X & X & X & NS & X & X & X & X & X & X & X & X & X & X & X & X \\ \hline
E.2.1.1.2: Disk-based  & ? & NS & X & NS & NS & NS & X & NS & NS & X & NS & NS? & NS? & X & NS & X & NS?\\ \hline
\rowcolor{myblue}
E.2.1.2: Node &  &  &  &  &  &  &  &  &  &  &  &  &  &  &  &  &  \\ \hline
E.2.1.2.1:User interface  & X & X & X & X & X & X & X & X & X & X & X & X & X & X & X & X & X \\ \hline
\rowcolor{myblue}
E.2.1.2.2: Resource management &  &  &  &  &  &  &  &  &  &  &  &  &  &  &  &  &  \\ \hline
E.2.1.2.2.1: Agent  & X & X & X & X & X & X & X & X & X & X & X & X & X & X & X & NS & NS \\ \hline	
E.2.1.2.2.2: Base resource  & X & X & X & X & X & X & X & X & X & X & X & X & X & X & X & X & X \\ \hline
\rowcolor{myblue}
E.2.1.3:  Cluster &  &  &  &  &  &  &  &  &  &  &  &  &  &  &  &  &  \\ \hline
E.2.1.3.1:  Configuration  & X & X & X & X & X & X & X & X & X & X & X & X & X & X & X & X & X \\ \hline
E.2.1.3.2:  Runtime  & X & X & X & X & X & X & X & X & X & X & X & X & X & X & X & X & X \\ \hline
\rowcolor{myblue}
E.2.2: Method  &  &  &  &  &  &  &  &  &  &  &  &  &  &  &  &  &  \\ \hline
E.2.2.1: Multicast  & X? & X & ? & NS? & X & X & X & ? & X & X & NS? & X & NS? & X & X & ? & X? \\ \hline
E.2.2.1.1: Atomic  & ? & ? & ? & NS? & X & X & ? & ? & X & ? & ? & ? & ? & ? & X & ? & ? \\ \hline
E.2.2.1.2: Virtual synchrony  & NS & NS & NS & NS & ? & X & NS & NS & X & NS & NS & NS & NS & NS & X & NS & NS \\ \hline
E.2.2.2: Broadcast  & ? & ? & ? & X & X & X & X & X & X & ? & X & X & X & X & X & X & X \\ \hline
E.2.2.3: Unicast  & ? & ? & ? & X & X & X & X & ? & X & X & X & X & ? & ? & X & ? & ? \\ \hline
E.2.2.4: IP socket  & NS & NS & NS & NS & NS & NS & NS & NS & NS & X & NS & NS & X & NS & NS & NS & NS \\ \hline
\rowcolor{myblue}
E.3: Resource management  &  &  &  &  &  &  &  &  &  &  &  &  &  &  &  &  &  \\ \hline
\rowcolor{myblue}
E.3.1: Type  &  &  &  &  &  &  &  &  &  &  &  &  &  &  &  &  &  \\ \hline
E.3.1.1: Base resource  & X & X & X & X & X & X & X & X & X & X & X & X & X & X & X & X & X \\ \hline
E.3.1.2: Application  & X & X & X & X & x & X & X & X & X & X & X & X & X & X & X & X & x \\ \hline
E.3.1.2.1: Agent-based  & X & X & X & X & X & X & X & X & X & X & X & X & X & X & X & NS & NS \\ \hline
E.3.1.2.1.1: Application (C- Siebel CRM, O-Oracle, S-SAP, T-Others W-WebSphere) & C, S, T & C, O, S, T & S, T, W & O, S & T$^{12}$ & O, S & O, S, T, W & S, T & O, S, W & O, S, T & T & O, S & S, W & C, O, S, W & O, S, W & S & C, O, S, T, W $^{15}$ \\ \hline
E.3.1.2.1.2: Database (A-SAP ASE, D-DB2, H-HANA, I-Informix, M-MySQL, O-Oracle, P-PostgreSQL, S-SQL Server, T-Teradata) & D, M, O, S & M, O & A, D, H, M, O, P, S & A, D, H, O, S & M $^{12}$ & D, M, O, P & D, H, O & O & A, D, M, O, P & A, D, I, M, O, P & O, M, P, S & A, D, H, O, P & A, D, I, M, O, P, S & A, M, P, O & D, H, I, M, O, P & D, H, O & A, D, M, O, P, S $^{15}$ \\ \hline
\rowcolor{myblue}
E.3.2: Method   &  &  &  &  &  &  &  &  &  &  &  &  &  &  &  &  &  \\ \hline
E.3.2.1: Policy-based   & X & X & X & X & X & X & X & X & X & NS & X & X & X & X & X & X & X \\ \hline
E.3.2.2: Rule-based   & ? & ? & ? & ? & ? & X & ? & ? & ? & X & ? & ? & ? & ? & ? & ? & X? \\ \hline
\rowcolor{myblue}
F:  Failure detection and recovery  &  &  &  &  &  &  &  &  &  &  &  &  &  &  &  &  &  \\ \hline
F.1: Monitoring   & X & X & X & X & X & X & X & X & X & X & X & X & X & X & X & X & X \\ \hline
F.1.1: Area   &  &  &  &  &  &  &  &  &  &  &  &  &  &  &  &  &  \\ \hline
F.1.1.1: Server   & X & X & X & X & X & X & X & X & X & X & X & X & X & X & X & X & X \\ \hline
F.1.1.2: Cluster   & X & X & X & X & X & X & X & X & X & X & X & X & X & X & X & X & X \\ \hline
F.1.1.3: Application   & X & X & X & X & X & X & X & X & X & X & X & X & X & X & X & X & X$^8$ \\ \hline
\rowcolor{myblue}
F.1.2: Type   &  &  &  &  &  &  &  &  &  &  &  &  &  &  &  &  &  \\ \hline
F.1.2.1: State-based   & X & X & X & X & X & X & X & X & X & X & X & X & X & X & X & X & X \\ \hline
F.1.2.2: Threshold-based   & ? & X & X & X & X & NS? & NS? & X & NS & X & NS? & ? & X & ? & NS? & NS? & X \\ \hline
\rowcolor{myblue}
F.1.3: Method   &  &  &  &  &  &  &  &  &  &  &  &  &  &  &  &  &  \\ \hline
F.1.3.1: Poll   & X & X & X & X & X & X & X & X & X & X & X & X & X & X & X & X & X \\ \hline
F.1.3.2: Push   & NS & NS & ? & NS & ? & NS & ? & ? & NS? & NS & NS & ? & NS? & NS & NS & NS? & ? \\ \hline
F.1.3.3: Event-based  & X$^2$ & NS? & NS? & X & ? & NS & NS & X & NS & NS & NS & NS & NS & NS & NS & NS & NS \\ \hline
\rowcolor{myblue}
F.2: Failover  &  &  &  &  &  &  &  &  &  &  &  &  &  &  &  &  &  \\ \hline
F.2.1: Reactive   & X & X & X & X & X & X & X & X & X & X & X & X & X & X & X & X & X \\ \hline
F.2.2: Proactive  & NS & NS & NS & NS & NS & NS & NS & NS & NS & NS & NS & NS & NS & NS & NS & NS & NS \\ \hline
\rowcolor{myblue}
F.3: Recovery level  &  &  &  &  &  &  &  &  &  &  &  &  &  &  &  &  &  \\ \hline
F.3.1: Resource  & X & X & X & X & X & X & X & X & X & ? & ? & X & X & X & X & X & X \\ \hline
F.3.2: Group   & X & X & X & X & X & X & X & X & X & ? & X & X & X & X & X & X & X \\ \hline
F.3.3: Node   & X & X? & X & X & X & X & X & X & X & X & X & X & X & X & X & X & X \\ \hline
F.4: Prediction  & NS & NS & NS & NS & NS & NS & NS & NS & NS & NS & NS & NS & NS & NS & NS & NS & NS \\ \hline
F.5: Simulation  & NS? & X & X & X & X & X & X & NS? & X & NS & NS? & X	& NS? & NS? & X & X & NS? \\ \hline
\rowcolor{myblue}
G: Consistency and integrity  &  &  &  &  &  &  &  &  &  &  &  &  &  &  &  &  &  \\ \hline
G.1: Cluster lock   & NS & ? & X & NS & X & NS & NS & X & X & ? & ? & X & X & NS & NS & X & NS \\ \hline
G.2: Quorum   & ? & X & NS & X & NS & X & X & X$^{20}$ & X & NS & NS & X & X & X & X & X & X \\ \hline
\rowcolor{myblue}
G.2.1: Device realisation  &  &  &  &  &  &  &  &  &  &  &  &  &  &  &  &  &  \\ \hline
G.2.1.1: Server   & ? & NS & NS & X & NS & X & NS & X & X & NS & NS & X & X & X & X & X & X \\ \hline
	\end{longtable*}
    }%
      \end{@twocolumnfalse}
  ]
 
\twocolumn[
\begin{@twocolumnfalse}
{
\scriptsize
	\begin{longtable*}[c]{|P{.152\textwidth}*{4}{|P{.01\textwidth}}|P{.01\textwidth}*{6}{|P{.01\textwidth}}|P{.01\textwidth}*{5}{|P{.01\textwidth}}|}
		\hline
		G.2.1.2: Disk   & ? & X & NS & NS & NS & NS? & X & NS & X & NS & NS & NS & NS & X & X & X & X \\ \hline
		G.2.1.3: File share   & NS & NS & NS & NS & NS & NS & X$^{21}$ & NS & NS & NS & NS & NS & NS & X & NS & X & X \\ \hline
\rowcolor{myblue}
G.2.2: Mode  &  &  &  &  &  &  &  &  &  &  &  &  &  &  &  &  &  \\ \hline
G.2.2.1: Server   & ? & NS & NS & X & NS & X & NS & X & X & NS & NS & X & X & X & X & X & X \\ \hline
G.2.2.2: Node   & NS & NS & NS & ? & NS & ? & ? & ? & ? & NS & NS & ? & ? & ? & ? & ? & X \\ \hline
G.2.2.3: Disk   & ? & X & NS & NS & NS & NS? & X & NS & X & NS & NS & NS & NS & X & X & X & X \\ \hline
G.2.2.4: File share   & NS & NS & NS & NS & NS & NS & X & NS & NS & NS & NS & NS & NS & X & NS & X & X \\ \hline
G.3: Dynamic quorum   & ? & ? & NS & X & NS & X & X & X & X & NS & NS & X & ? & X & X & X & X \\ \hline
\rowcolor{myblue}
G.4: Isolation   &  &  &  &  &  &  &  &  &  &  &  &  &  &  &  &  &  \\ \hline
G.4.1: Fencing  & NS & X & NS & X & NS & X & X & X & X & X & NS & X & X & X & X & ? & NS \\ \hline
G.4.1.1: Resource  & NS & NS? & NS & X & NS & X & X & X & X & X & NS & X & X & X & X & ? & NS? \\ \hline
G.4.1.2: Node  & NS & X & NS & X & NS & X & ? & ? & X & X & X$^{19}$ & ? & X & ? & X & ? & NS \\ \hline
G.4.2: Shutdown  & NS & X & X & NS & X & x$^{13}$ & X & X & X & X & NS & X & X & X? & NS? & X & NS \\ \hline
\rowcolor{myblue}
H: Data synchronization  &  &  &  &  &  &  &  &  &  &  &  &  &  &  &  &  &  \\ \hline
H.1: Shared storage  & X & X & X & X & X & X & X & X & X & X & ? & X & X & X & X & X & X \\ \hline
H.2: Shared-nothing  & X & X & X & X & X & X & X & X & X & X & X & X & X & X & X & X & X \\ \hline
H.2.1: Replication  & X & X & x & X & X & X & X & X & X & X & X & X & X & X & X & X & x \\ \hline
\rowcolor{myblue}
H.2.1.1: Type  &  &  &  &  &  &  &  &  &  &  &  &  &  &  &  &  &  \\ \hline
H.2.1.1.1: Application-based  & X$^6$ & X$^6$ & X$^6$ & X$^6$ & ? & X$^6$ & X$^6$ & X$^6$ & X$^6$ & X$^6$ & X$^6$ & X$^6$ & X$^6$ & X$^6$ & X$^6$ & X$^6$ & X$^6$ \\ \hline
H.2.1.1.2: Array-based  & x & x & x & x & x & x & x & x & x & ? & ? & x & x & x & x & ? & x \\ \hline
H.2.1.1.3: Cluster-based  & NS & NS & X & NS & NS & NS & NS & NS & NS & NS & X & NS & X$^{18}$ & NS & NS & NS & NS \\ \hline
H.2.1.1.4: Host-based  & x$^{17}$ & x & x & X & x & X & X & X & X & ? & NS & X & X & X & X & ? & x \\ \hline
\rowcolor{myblue}
H.2.1.2: Method  &  &  &  &  &  &  &  &  &  &  &  &  &  &  &  &  &  \\ \hline
H.2.1.2.1: Synchronous  & X & X & X & X & X & X & X & X & X & ? & X & X & X & X & X & X & X \\ \hline
H.2.1.2.2: Asynchronous  & X & X & X & X & X & X & X & X & X & ? & X & X & X & X & X & X & X \\ \hline
H.2.2: Mirroring  & x & x & x & x & ? & x & x & x & x & x & x & x & x & x & x & ? & x \\ \hline
H.2.2.1: Synchronous  & x & x & x & x & ? & x & x & x & x & x & x & x & x & x & x & ? & x \\ \hline
H.2.2.2: Asynchronous & x & x & x & x & ? & x & x & x & x & x & x & x & x & x & x & ? & x \\ \hline \\ 
\multicolumn{18}{l}{Key: ? -- No information; X -- Supported; 
    	NS -- Not supported;
    	X? -- Supported (not explicitly stated in the documentation, but this}\\
    	\multicolumn{18}{l}{interpretation has been made by analysing the documentation);
    	x -- Supported together with additional components, and an example}\\
    	\multicolumn{18}{l}{is replication support by the operating system volume manager;
    	NS? -- Not supported (not explicitly stated in the documentation,}\\ \multicolumn{18}{l}{but this interpretation has been made by analysing the documentation.}
	\end{longtable*}
    \begin{description}
     \scriptsize
    	\item $^{1}$ Supported only on virtualized environments.
    	\item $^{2}$ Intelligent monitoring framework.
    	\item $^{3}$ Replication or mirroring support by additional tools is included.
    	\item $^{4}$ Supported on both virtual and physical environments.
    	\item $^{5}$ 64 nodes are supported for the hub, while leaf nodes can support many more.
    	\item $^{6}$ Replication is provided natively by an application, but a HAC must support the feature.
    	\item $^{7}$ If an application supports parallel deployments.
    	\item $^{8}$ OpenSAF provides a generic development package; it can be ported to other UNIX and Linux flavours.
    	\item $^{9}$ LPARs: 2 logical partitions (LPARs) on IBM PowerVM
    	\item $^{10}$ Supported on System x hardware that is based on the x86 platform.
    	\item $^{11}$ Supported only on Oracle’s x86 platforms.
    	\item $^{12}$ The implementer can develop application support.
    	\item $^{13}$ Fencing by STONITH (Shoot the Other Node in the Head).
    	\item $^{14}$ The maximum number of nodes on Linux is 32, and, for AIX, it is 130.
    	\item $^{15}$ Application vendors provide application support for WSFC.
    	\item $^{16}$ Usually, guest heartbeat is passed to a host.
    	\item $^{17}$ Replication features of a virtual machine can also be used.
    	\item $^{18}$ Replication feature is provided by the product DataKeeper, which is part of the SIOS Protection Suite.
    	\item $^{19}$ Fencing as a concept is not employed, but, instead, the node with the problem is put into a waiting state..
    	\item $^{20}$ The solution uses a quorum technique called cluster integrity.
    	\item $^{21}$ Implies repository disk.
    	\item $^{22}$ Supported by SafeKit 7.1.3.
    	\item $^{23}$ The vendor provided most details.
    	\item $^{24}$ Version for Linux. Current version for HP-UX is A.11.20.
    	\item $^{25}$ The maximum number of supported nodes for Linux is 32, while for HP-UX, it is 16.
    	\item $^{26}$ The maximum number of supported nodes on Solaris on x86 is 8, and Solaris on SPARC supports 16.
    	\item $^{27}$ None of the surveyed HACs support container-based technologies.
    	\item Note:
Although the operating system version is not stated, it is the most recent version at the time this survey was carried out (The survey was conducted mainly between October 2018 and June 2019. The survey was updated during July 2021, September 2021 and then October 2021).\\ 
    \end{description}
    }%
     \end{@twocolumnfalse}
  ]

\noindent

\begin{table*}
\sffamily
  \centering\scriptsize
		\caption{The surveyed HACs, versions and vendors}
		\label{tab:surveyed HACs-versions-vendors}
		 \setlength{\tabcolsep}{4pt}
  \begin{tabular}{P{.28\textwidth}P{.085\textwidth}P{.425\textwidth}P{.13\textwidth}}
        \toprule \textbf{Surveyed HAC}  & \textbf{Vendor} & \textbf{Surveyed HAC}  & \textbf{Vendor}\\ \midrule
        ApplicationHA 6.2* & Veritas &  RSF-1 3.9.10* & High-Availability\\
        ClusterLabs stack 2.3.2* & ClusterLabs &  SafeKit 7.2* & Evidian \\ 
        Clusterware 12c* & Oracle & Serviceguard A.12.20* & HPE\\ 
        EXPRESSCLUSTER X 3.3* & NEC &  SIOS Protection Suite 9.2* & SIOS\\\ 
        InfoScale Availability 7.3.1* & Veritas & Solaris Cluster 4 & Oracle\\
        OpenSAF 5.17.07* & SA Forum & SUSE Linux Enterprise High Availability Extension 12* & SUSE\\ 
        PowerHA SystemMirror 7.2.1  & IBM & Tivoli System Automation for Multiplatforms (SA MP) 4.1*  & IBM\\ 
        PRIMECLUSTER 4.5* & Fujitsu & Windows Server Failover Clustering (WSFC) 2016 & Microsoft\\
        Red Hat High Availability Add-On 7.0*& Red hat\\ 
              \bottomrule
              \multicolumn{4}{l}{* = solution that functions as middleware}
      \end{tabular}
       \end{table*}

 Each of these HACs functions as middleware, which means that it creates an additional layer on the top of an operating environment. The HACs from the second group, which comprises the remaining three solutions, are tightly integrated with an operating system and make use of the features that are offered by an operating environment. The latter type of HAC functions as part of an operating environment, operating in the kernel mode and directly interacting with operating system functionalities. While such features can make a HAC more efficient, they may also create problems with modularity and portability, and that is why, for example, such HAC solutions only support specific operating systems. Furthermore, the lifecycle management of such a HAC solution also becomes the operating system's lifecycle management. WSFC has already embraced this approach, and it is entirely integrated with the Windows server operating system.

Major software and hardware vendors have their HAC solutions. However, some of them are supported only by the operating environment and platform from the vendor. An example of this is WFCS, which is only available on the Windows server enterprise edition.
On the other hand, some independent vendors specialise in HAC products, and these vendors can support multiple operating systems and platform combinations. Typically, such HACs belong to the middleware group. 

Cloud deployment has also come to play an important role. In the early days of cloud computing, a separate development of HAC was considered. This led to the development of specific HAC solutions, such as ApplicationHA by Veritas and vSphere App HA by VMware. However, a better approach is to port existing solutions to the cloud environment, which made developing cloud-specific HAC solutions unnecessary. An such, solutions like App HA by VMware were discontinued. However, HACs in the public cloud comes with limitations. For example, using shared storage is a challenge.
On the other hand, this has contributed to developing enhancements to enable deploying HACs in the cloud.  One such enhancement is the so-called \textit{storage-less} or \textit{SANless} HAC, which allows HACs to operate without shared storage. Moreover, the transition to cloud services models, such as SaaS, PaaS, and IaaS, changes the way HACs are deployed and managed. Likewise, roles and responsibilities for managing a HAC with the different service models also change.

Furthermore, the introduction of \textit{multi-clouds} can also complicate a HAC deployment, not least from a roles and responsibilities perspective. Somasekaram highlights the issues with roles and responsibilities of HA and DR solutions in the context of outsourcing \cite{mastersthesis}. He argues that the issues are valid even for the cloud environment because the cloud is regarded as outsourcing, and cloud providers are usually responsible for multiple layers (e.g., network and storage). At the same time, other suppliers manage the rest of the layers.

Similar to the challenges described for the deployment environment public cloud and host virtual, the emerging deployment environments fog and edge also face challenges \cite{singh2021fog,yousefpour2019all}. When used with the host virtual, the challenges are the same as the public cloud. On the other hand, when the host container is used in all deployment environments, ensuring high availability for stateful applications hosted in containers becomes a challenge. Containers run as a process in user space, and this may restrict the implementation of HAC features that require running in kernel space \cite{ramos2021machine}. Moreover, containers typically support a single application or a service in a container which means a HAC cannot deploy agents in the same container to manage the application resources \cite{VeritasTechnologiesLLC2020}. To overcome this limitation, the commonly implemented container orchestration system Kubernetes provides a sidecar option (i.e., a separate container), which enables deployments of HAC-related components there. The sidecar container runs along with the container that hosts the application. Using this approach, commercial vendors started proving HACs for containers. InfoScale availability (formerly Cluster Server—VCS) for containers from Veritas is one such HAC for Kubernetes. This HAC provides monitoring, integrated I/O fencing, arbitration and shared storage using a Container Storage Interface (CSI) plugin \cite{VeritasTechnologiesLLC2020} to ensure that the HAC can deliver HA for the application. The solution requires at least two private networks to enable cluster communication and one public network to facilitate heartbeat communication. There are also research projects that explore the use of existing HACs such as Pacemaker/Corosync~\cite{vayghan2019microservice} and OpenSAF~\cite{alahmad2018high} to support container-based applications. 

\textit{Latency} over long distances has traditionally been a major problem for HACs. However, the technology has evolved and techniques are currently available to reduce latency considerably, enabling the setting up of HACs across substantial distances. \textit{Atomic broadcast} and \textit{multicast} (total order messaging) are often associated with fault-tolerance in distributed systems; hence, there are persuasive arguments to employ it even for HAC communication \cite{ServiceAvailabilityForum2011}. However, it is only employed by some of the HACs today. 

On the whole, \textit{prediction} is absent from the surveyed HAC solutions. Most solutions employ a poll-based monitoring mechanism that is often state-based, meaning that only the states of the resources are monitored. Moving towards \textit{industry-standards} has also been observed in some areas, such as when using the IPMI to shut down nodes as part of isolating a problematical node. The \textit{SCSI-3} interface is widely employed to isolate on a resource level, and often as part of \textit{fencing}. The \textit{quorum} concept is commonly employed so that a cluster can take action upon a situation that leads to the partitioning of a cluster.

In conclusion, the current HAC solutions for EAs are dominated by commercial vendors (15 out of the 17 surveyed solutions). This is unsurprising because customers look for HAC solutions for their business-critical applications, and, as such, proper support is paramount. However, this also means that the vendors conduct most of the research. There are, however, some open-source initiatives, and two active initiatives are OpenSAF and ClusterLabs stack (Pacemaker/Corosync). The open-source initiatives often focus on Linux, and there have been different projects to develop a consistent HAC solution for Linux. While such efforts have been split into other projects or discontinued, some of the components are still active, and the current open-source cluster solutions are a combination of various initiatives.
The main components of the current setup of the ClusterLabs stack are Corosync, Pacemaker, DRBD, STONITH, and a diverse range of  application agents, which are packaged under a ClusterLabs stack. OpenAIS was an initiative to support implementing Application Interface Specification (AIS) developed by the Service Availability Forum (SA Forum), and Corosync originated from that initiative. Pacemaker is a cluster resource manager (CRM) tool that originates from the Linux-HA project. 

Application agents follow the standard API established by the Open Cluster Framework (OCF), which helps standardise the application resource management. Both SuSe Linux Enterprise HA Extension and Red Hat Enterprise Linux HA add-on use Pacemaker, Corosync, the OCF concept and many other open-source components. OpenSAF, on the other hand, focuses on the telecommunications sector, where there is a need to support very high availability for carrier-grade servers that operate in the telecommunication infrastructure. However, there have been initiatives to deploy OpenSAF in a range of environments, such as the cloud. For example, Kanso et al. \cite{kanso2013achieving} proposed an OpenSAF based deployment in the cloud, but it too focuses on telecom applications.  The challenge with open-source initiatives is to secure proper support, which is crucial for EAs.
On the other hand, Red Hat and SuSe provide such support even though they have developed their HACs using mainly open-source components. It must be noted that there have been several projects related to the development of HACs, both commercial and open-source, over the years. However, many of them are no longer active, and examples include FailSafe by Silicon Graphics (SG) in the commercial area, while HA-OSCAR represents an open-source equivalent.

\section{Future Directions}
\label{sec:future-directions-taxonomy-survey}

We have identified several limitations, challenges and opportunities as part of constructing the HAC taxonomy from Section~\ref{sec:taxonomy} and conducting the survey from Section~\ref{sec:survey}. The limitations and challenges are from an implementation perspective and an operations viewpoint, while opportunities can improve the overall HAC solutions. Using the identified limitations, challenges and opportunities, we discuss future research directions.

\subsection{Limitations \label{subsec:limitations}}
The HAC limitations presented in this section apply to a majority of the HAC solutions that we have studied, with limitations L1, L5, and L7 common for all solutions. 

\textbf{L1. Standardisation}. Standardisation of the HA domain, its components, and related processes is missing. For example, the terminology used by HAC solutions differs considerably. Standardisation could improve research approaches and could enable better discussions and research quality. Furthermore, the lack of standardisation makes it challenging to develop standard APIs that can function with multiple solutions to support specific functionalities, for instance, application-specific agent development. We have addressed this lack of standard terminology using consistent terminology while constructing the taxonomy and performing the subsequent survey.

\textbf{L2. Virtual environments}. The separation between host and guest in virtualised environments complicates some of the functionalities of HACs, such as coordinated monitoring of two operating environments, guest and host, which must be correlated when hosting a critical application. If such a setup is not in place, a guest HAC may not be aware of the host at all. If there are problems in the host which impact all the guests hosted there, the guest HAC may not be able to recognise the problems \cite{loveland2008leveraging}, which could potentially impact the application. Likewise, if the guest application experiences problems, the host may not react since it is unaware of any issues except when hardware resource utilisation significantly increases. Kanso et al. \cite{kanso2013achieving} highlighted the problem with a guest HAC that is not aware of the host environment. Some HAC solutions promote a solution by running additional components on the host that also interact with the guest HAC. However, there is no uniformity for deploying such components because they may differ based on the virtual environment, such as VMWare or kernel-based VM (KVM). In KVM, additional tools are typically required on the host, while VMware comes with a set of accessories that can be used instead so that no additional means are required. Some HAC solutions, such as ApplicationHA on KVM, employ a separate HAC installation on a host machine. For example, ApplicationHA on the guest can interact with the host HAC. This setup can support monitoring of the host and enable the use of features that are not otherwise available in the guest environment due to restrictions. However, a heterogeneous virtual environment with different operating environments for hosts and guests may also complicate the cross-deployment of a HAC as each operating system, platform, and virtual environment comes with restrictions.

\textbf{L3. Cloud environment limitations}. Both private and public clouds come with limitations. In such a cloud environment, particularly in an IaaS model, customers have access to a guest environment (e.g., VMs). To support an EA, a HAC will require access to some host elements well. In addition, the host environment must be monitored as well as part of a holistic HAC approach, which may mean deploying additional tools, as described in \textbf{L2}, on the host. The limitations of a cloud environment may require changes in the architecture of the HAC, hence also the protected application \cite{Nabi2016AvailabilityArt}.

\textbf{L4. Public cloud limitations}. In addition to what is described in \textbf{L2} and \textbf{L3}, the public cloud has some additional infrastructure-related restrictions, which are usually different from those of a private cloud. For example, shared storage is not typically supported \cite{AmazonWebServicesInc2016}. Hence any shared-storage-based HAC must find an alternative solution that implies that shared-disk-based quorums cannot be employed. Further, there could be additional restrictions impacting the core functionality of a HAC, such as multicast or broadcast communication not being allowed \cite{AmazonWebServicesInc2016}, which would impact the HAC's ability to communicate. Again, this means alternative solutions must be identified and implemented by adding new tools and procedures, which may, in turn, add more complexity to a solution. Moreover, if an application is deployed in a virtual environment, additional restrictions, described in \textbf{L2}, apply. For instance, the deployment of additional HAC tools on a host, as explained in \textbf{L2}, is usually not possible as hosts are managed entirely by cloud providers in such settings (\textbf{L3}).

\textbf{L5. Rating of errors}. Often a severity rating is not used for errors on a resource level, which means that all errors are treated equally. Adding severity levels would help distinguish between the different types of errors and by the different modules of HACs (e.g., monitoring, failure management) so that actions can be taken accordingly. In addition, multilevel severity would help to improve the recovery process so that, in some cases, errors can be disregarded, indicating that such errors do not result in a complete failover.

\textbf{L6. Standardisation of error, failure, and event message representation}. The current approach is very much individualised to different HAC solutions, implying no standard structure for log messages. This makes it hard to develop a general solution to analyse log messages (e.g., for analytical purposes). Furthermore, several modules of a HAC (e.g., monitoring, failure management) may write the same error messages with the same timestamp when a resource fails, making it challenging to mine the log for distinct error messages. Moreover, log sources can also vary as some HAC solutions may employ more than one log source. The difference presents a challenge in mining data from log sources, as it will require one or more data extraction interfaces for each HAC solution.

\textbf{L7. Rating of resource and resource group dependencies}. Resource and resource group dependencies are not always rated, which means that the same failover and recovery policies are applied to all dependencies, regardless of the strength of the dependencies. The dependency rating describes how the failure of a resource can impact another resource through a dependency connection and on what level, which can ultimately influence the mitigation action. 

\textbf{L8. Application monitoring}. Even though many of the current HAC solutions employ application monitoring, application-spe\-cif\-ic errors (e.g., hang situations), are not usually captured. Furthermore, application-related errors are often difficult to monitor with a HAC. This may require additional modules and steps, such as logging in to an application, to detect such failure. Hence, the current situation is that an application may be completely unresponsive, yet it is still regarded as running by a HAC. Therefore, such errors do not trigger any action until the problem is reported by the users of the application. Consequently, this will also result in incorrect values for MTTR and MTBF since no accurate time of failure is available, thus providing unreliable figures for availability.

\subsection{Challenges}
Challenges are associated with functionalities or features that can be implemented to improve the effectiveness of HACs, but that are difficult to realise due to limitations and other constraints. 

\textbf{C1. Roles and responsibilities}. HACs work closely with operating environments and infrastructure components to provide the required HAC functionalities, such as heartbeat, monitoring, fencing, and quorum. While the roles and responsibilities of the experts in charge of setting up a HAC change with the different cloud service models \cite{Vacca2016CloudChallenges}, it is unlikely that one team can manage a complete HAC implementation and operations. Instead, multiple teams and even organisations must work together to support HAC implementation and operations. Moreover, a heterogeneous virtual environment further complicates the setup because at least two operating systems will be associated with host and guest, which means different teams are usually designated to support the host and guest environments. This means that there must be a support process that links all the different teams together according to a well-defined roles and responsibilities matrix. Moreover, the related support processes, for instance, change and incident management, must be designed accordingly. 

\textbf{C2. Lifecycle management}. The combination of many application agents, HAC components, VMs, operating systems, and platforms complicates the lifecycle management of HACs. While having a standard across architecture components (including agents) can reduce the number of combinations, this is extremely difficult to achieve. In particular, in virtual environments, lifecycle management must take into account other elements, such as host and guest operating environments on various VMs, which adds further complexity, as described in \textbf{L2}. The number of combinations may prompt more threads of lifecycle management. For example, when an EA vendor releases an update to the application, a HAC vendor must also make sure to release an update of the HAC or agent to support the changes in the application. 

\textbf{C3. Client-state synchronisation}. Client-state synchronisation for EAs is a difficulty. However, if achieved by a HAC, it can improve availability significantly because it can transfer user sessions in the event of a failover, which means that no user data is lost. When a failover takes place, all user input that is not saved is lost. When the failover is complete, users can log in again to establish new sessions and start their work from scratch. If an EA supports thousands of users, this means losing countless hours of work.
On the other hand, if a client-state synchronisation can be achieved, it will preserve all connections and sessions, saving considerable time. It is also likely that, with faster failovers using client-state synchronisation, users will not even notice that a failover has taken place. Instead, they will be able to continue working as if nothing has happened. However, state synchronisation for an EA is a significant challenge because it requires replicating user connections, user sessions, user context, session context, user work, and global and local variables. While solutions with a limited scope, such as a firewall, widely employ client-state synchronisation, these are difficult to adopt for the much more complex settings of EAs. Since the problem is about preserving user sessions and related data, in many cases, an applications server layer may also need to be synchronised, as they are the front-ends for user communication in a multi-tier system.
Furthermore, applications with a sizable workload require substantial time to stop and start application components in a specific order. Some portion of that time is consumed on ending user sessions gracefully during the stop and establishing non-user (e.g., batch) sessions at the start. However, client-state synchronisation may reduce that time significantly since user data would be already synchronised across the HAC members. 

\subsection{Opportunities}
We have identified a set of opportunities that can improve HAC solutions considerably, typically by overcoming HAC limitations from Section~\ref{subsec:limitations}. For instance, introducing probabilistic and statistical methods as detailed in opportunity \textbf{O5} below requires that ratings of errors and dependencies are in place. Hence, the exploitation of opportunity \textbf{O5} requires solutions to limitations \textbf{L5} and \textbf{L7}.

\textbf{O1. Architecture components}. HAC solutions employ different architecture components, and therefore having a standard and modular architecture will help standardise these components. This assumes that such an architecture will consist of standard modules, and that a HAC solution can choose to implement only a subset of modules, but it can always refer such modules to the standard modules. A set of specifications can support defining the roles of the modules and even provide means to develop interfaces (e.g., APIs). Solutions that are developed using the APIs can potentially be used with multiple HAC solutions. Moreover, the approach would aid in simplifying and interpreting architecture components while enabling the development of approaches for new and emerging technologies (e.g., containerisation), standard testing, and benchmarking. 

\textbf{O2. Evaluation of historical data}. HACs produce a large volume of data, and such data can be invaluable when analysing past events and mitigations. These data are generated mainly through logging of events, failures, recoveries, and failovers. Historical data, together with current data, can be analysed to identify patterns, enabling proactive approaches to ensuring high availability. Therefore, evaluation of past data and current data can be used to predict failures of a repeating nature and other related failures. 

\textbf{O3. Reliable cluster communication protocols}. The reliability of cluster communication can be increased significantly by employing protocols with atomic features such as TOTEM \cite{dake2008corosync}. These features are only supported by a few HAC solutions today. Employing a standard protocol will also enhance development in the areas, as more people can be involved in the development, which means that issues can be addressed quickly. 

\textbf{O4. Monitoring}. Most HAC solutions use a poll-based monitoring method, which is linked to performance problems \cite{VeritasTechnologiesLLC2017}. If the polling frequency increases, it will improve the monitoring data quality because more up-to-date data will then be available. However, there is an additional overhead associated with frequent polling of many resources, which could be resource-demanding. Furthermore, detecting application-specific errors might also present some challenges as described in \textbf{L8}. The monitoring functionality of a HAC may not detect such cases since HACs often focus on monitoring state changes of a resource or a resource group. Therefore, relevant monitoring models should be evaluated to improve the data quality while reducing the performance overhead. The current monitoring type is mostly state-based. However, a different option might be to use a standard API to interact with the operating environments so that the enhanced monitoring features of the operating systems can be utilised. Though this approach may still require an application-specific development, it can be simplified by using standard APIs, as discussed in \textbf{L1} and \textbf{O1}.

\textbf{O5. Incorporation of probabilistic and statistical methods}.  Such methods are not employed currently, but they can improve effectiveness significantly and reduce downtime by analysing data, checking behaviours and providing predictions. In addition, such methods will also improve the quality of the service for HACs and their components, in general, and promote a more robust proactive approach than currently employed by mostly reactive mechanisms. One example of such an improvement is introducing statistical analysis to enable the management of quorum services more intelligently. 

\textbf{O6. Analytical services}. Analytical services will help identify patterns in the behaviour of HACs and their components while also providing a consolidated view of total downtime and causes. Analytical services can also incorporate data from multiple sources so that data can be combined to provide reliable analysis and even produce predictions on potential failures. An example is that if some HAC components manifest intermittent failures before complete failure, patterns can be analysed to estimate the subsequent failure or an eventual complete failure. 

\textbf{O7. Benchmark}. A standard benchmarking approach that can measure availability at a granular level will improve the performance measurements of HACs, while also enabling more a natural comparison between different solutions. 

\textbf{O8. Security}. \textit{HAC security} is a rarely concern. However, unauthorised access to the services of a HAC means effectively that the protected application is also jeopardised because a HAC has typically complete control of the operations of the protected application. Security is of particulat concern in cloud environments with shared responsibilities (\textbf{C1}), since multiple teams assume responsibility for the different layers, which may present new vulnerabilities without a proper security model in place. Moreover, operating a HAC solution in a public cloud may also introduce new vulnerabilities \cite{Chow2009ControllingCloud}, mainly when new and alternative solutions must be introduced due to restrictions, as described in \textbf{L4}.

\section{Conclusions}
\label{sec:summary-conclusion-taxonomy-survey}

In this article, we presented a comprehensive taxonomy and a two-part survey of high-availability clusters. The first part of the survey, delivered while describing the elements of the taxonomy, provides an overview of the HAC research landscape. The second part employs the taxonomy to survey end-to-end HAC solutions developed to support enterprise applications. Finally, we detailed HAC limitations, challenges and opportunities identified while constructing the taxonomy and conducting the two-part survey. Using these, we discuss future research directions for high-availability clusters. In particular, an adaption of fully functional HACs for cloud-deployed enterprise applications can significantly improve the availability of these applications. Similarly, exploiting historical data through the use of probabilistic approaches to predicting future failures and other relevant events can improve the effectiveness of HACs. Last but not least, HAC support for client-state synchronisation has the potential to deliver zero downtime for an important range of failures affecting enterprise applications.




 \bibliographystyle{elsarticle-num} 
 \bibliography{reference}





\end{document}